\documentclass[aps,twocolumn,pra,tightenlines,floatfix,showpacs]{revtex4}
\usepackage[dvips]{graphicx}
\usepackage[english]{babel}
\usepackage{amsmath}
\usepackage{amssymb}
\usepackage{times}
\usepackage{bm,times}
\newcommand{\bs}{\begin{split}}
\newcommand{\es}{\end{split}}
\usepackage[dvips]{graphicx}

\newcommand{\mb}[1]{{\mathbf{#1}}}
\newcommand{\be}{\begin{equation}}
\newcommand{\ee}{\end{equation}}
\newcommand{\ba}{\begin{eqnarray}}
\newcommand{\ea}{\end{eqnarray}}
\newcommand{\ek}{\epsilon_{\mathbf{k}}}
\newcommand{\Ek}{E_{\mathbf{k}}}

\newcommand{\uksq}{{u_{\mathbf{k}}^2}}

\newcommand{\vk}{{v^{}_{\mathbf{k}}}}

\newcommand{\vksq}{{v_{\mathbf{k}}^2}}

\newcommand{\Eku}{E_{\mathbf{k},\uparrow}}
\newcommand{\Ekd}{E_{\mathbf{k},\downarrow}}
\newcommand{\xikk}{\xi_{\mathbf{k},3}}
\newcommand{\xik}{\xi_{\mathbf{k}}}
\newcommand{\sumk}{\sum_{\mathbf{k}}}

\renewcommand{\d}{\mathrm{d}}

\newcommand{\p}{\partial}

\def\ket#1{|#1\rangle}

\begin{document}

\title{Theory of Radio Frequency Spectroscopy
Experiments in Ultracold Fermi Gases and Their Relation to
Photoemission Experiments in
the Cuprates}

\author{$^{1,2}$Qijin Chen, $^1$Yan He,
$^{1}$Chih-Chun Chien, and $^1$K. Levin}

\affiliation{$^1$James Franck Institute and Department of Physics,
University of Chicago, Chicago, Illinois 60637, USA}
\affiliation{$^2$Zhejiang Institute of Modern Physics and Department of
Physics, Zhejiang University, Hangzhou, Zhejiang 310027, China}

\date{\today}

\begin{abstract}
In this paper we present an overview
of radio frequency (RF) spectroscopy in the atomic Fermi superfluids.
An ultimate goal
is to suggest new directions in
the cold gas research agenda from the condensed matter
perspective.
Our 
focus is on the experimental and theoretical
literature of cold gases
and photoemission spectroscopy
of the cuprates particularly as it
pertains to areas of overlap.
In addition to a comparison with the cuprates, this paper contains
a systematic overview of the theory of RF spectroscopy, both
momentum integrated and momentum resolved. It should be noted
that the integrated and momentum resolved forms of photoemission are 
equally important in the high $T_c$ cuprate literature.
We discuss
the effects of traps, population imbalance, final state interactions
over the entire range of
temperatures and compare theory and experiment, most notably in the context
of recent tomographic scans in population imbalanced gases.
We show that this broad range of phenomena can be accomodated
within the BCS-Leggett description of BCS-BEC crossover and that this scheme
also captures some of the central 
observations
in photoemission
experiments in the cuprates. In this last context, 
we note that the key themes which have emerged in cuprate photoemission
studies involve characterization of the fermionic self energy, of
the pseudogap and of the effects of superconducting coherence (in passing
from above to below the superfluid transition temperature,  $T_c$).
These issues have a counterpart in the cold Fermi gases
and it would be most useful in future to use these atomic systems to address these
and the more sweeping question of how to describe that anomalous superfluid 
phase which forms in the presence of a normal state excitation gap.
\end{abstract}

%Eq.~(\ref{eq:STOP})

%Appendix \ref{App:STOP}
%Section \ref{sec:STOP})
\maketitle

\section{Introduction and Motivation}
\label{sec:1}
There is considerable excitement surrounding the
discovery
\cite{Jin3,Grimm,Jin4,Ketterle3,KetterleV,
Thomas2,Grimm3,ThermoScience,Salomon3,Hulet4}
of superfluidity
in the ultracold Fermi gases. What is novel about these new superfluids is
that one can tune the attractive interaction from weak (as in the BCS
limit) to strong as in the Bose Einstein condensation (BEC) regime.
These experiments will continue to impact condensed matter
physics by providing, at the least, a new class of ``materials" which elucidate a
very powerful generalization of BCS theory. A number of people
\cite{ourreview,Varenna,Strinaticuprates,randeriareview} have
also argued that this BCS-BEC crossover might be relevant to the cuprate
superconductors. Because of their anomalously short coherence length
it is claimed \cite{LeggettNature}
that these materials are ``mid-way between BCS and BEC".
That is, the attractive interaction driving the superconducting
pairing may be stronger than that in conventional superconductors.
In this way the tuneability of the interaction strength in
the Fermi gases provides an ideal model system with which to study the physics
of the short coherence length cuprates and the role of
strong attraction (generally associated with high transition
temperatures).
From a very different perspective, it
has also been argued that in future optical lattice experiments
\cite{GeorgesReview}
involving the atomic Fermi gases, one will be able to simulate repulsive
Hubbard models and thereby investigate the ``Mott physics" aspects
\cite{LeeReview} of high $T_c$ superconductivity.

While condensed matter physicists have a wealth of well-developed techniques
for characterizing electronic superconductors, the tools currently available
to the atomic physicists who study the Fermi gases are far more
limited. Moreover, it is not at all straightforward to determine something
as commonplace as the temperature in the gas, although some impressive
progress \cite{ThermoScience,ThomasUnitary,KetterleVarenna}
has been made along these lines.

This paper is devoted to
addressing one of the most powerful
techiques currently being applied to the Fermi gases: radio frequency
(RF) spectroscopy. We will show how this technique is similar to that of
photoemission in condensed matter
physics and exploit the analogy, already discussed in
the literature \cite{Jin6}, between momentum resolved
RF and angle resolved photoemission spectroscopy (ARPES). As a background
for both communities, we
review some of the experimental and theoretical
literature on RF spectroscopy (of cold gases)
and photoemission spectroscopy
(of the cuprates).  We argue that there are
a number of issues which have been central to high temperature superconductivity
which would be useful to address more systematically  in the ultracold Fermi gases.
Perhaps the most notable example of commonality \cite{ourreview,Varenna}
in this regard
is the ubiquitous pseudogap phase which is at the core of
current studies in the high $T_c$
superconductors and has emerged as important in the ultracold Fermi gases.

We begin by focusing on the overlap of the experimental
concepts behind photoemission experiments \cite{arpesstanford_review,arpesanl}
in the cuprates and RF spectroscopy in the atomic Fermi gases.
We will see that both experiments reflect the behavior of the
all important fermionic spectral function $A ( \bf{k}, \omega)$ which
characterizes completely the single fermion or one-particle properties of a
given many body system.
In simplistic terms, the driving force motivating the photoemission
studies in the cuprates
is to acquire an understanding of the ``mechanisms" and nature of
superconductivity. There has been a recent emphasis on high temperatures
near $T^*$, where the pseudogap turns on and on the region from above to
below the superfluid transition temperature, $T_c$.
By contrast in the ultracold gases, the RF spectra have been
used to characterize the pairing gap $\Delta$-- much like tunneling is used
in conventional superconductors.
There has been a recent emphasis on very low temperatures $T << T_c$
and in particular in quantifying the size of $\Delta$ at $T=0$.

Some of the key issues which have emerged in photoemission studies
of the
cuprates
involve (i) a characterization of the self energy contained in
the spectral function.
Different empirical models \cite{Pepin} have been deduced which, it is argued,
might ultimately hold the clue
as to the nature of the mediating boson.
(ii) Also important is the origin of the all important pseudogap.
There is a debate \cite{LeeReview,ourreview} about whether
this gap is a signature of a hidden order parameter or whether it reflects
the incipient pairing which ultimately leads to the condensed phase at
lower $T$.
(iii) It is viewed as extremely important
to arrive at an understanding of how
superconducting coherence manifests itself in this spectroscopic
experiments as one goes from the
normal to the ordered phase. This is a complicated question, given
the presence of a normal state (pseudo)gap.
Finally, other issues of interest are the nature of the order parameter
and pseudogap symmetry
(which have been shown to be consistent with $d$-wave).

In the cold gases
an underlying goal has been
to test different theories of BCS-BEC crossover, particularly
establishing the most suitable ground state and its quantitative
implications such as the pair size \cite{Ketterlepairsize}.
The parameters which quantify
the nature of
the scale-free or ``unitary" gas have also been addressed.
Of additional interest are studies on how population imbalance
\cite{Rice1,Rice2,MITPRL06,ZSSK06}
can co-exist with
superfluidity. Here new phases associated with, for example,
the exotic \cite{FFLO} Larkin- Ovchinnikov, Fulde-Ferrell (LOFF) form
of pairing  have been contemplated.
Even more topical is
the behavior in the limit of extreme imbalance
\cite{MITPRL06,ZSSK06}.

One can see that, despite the similarities in
these two spectroscopic techniques, the research agenda in the two communities is
rather different. In the high temperature superconductors,
the focus has been
around the temperature regime near $T_c$.
Furthermore, quantitative issues are viewed as of considerably
less importance than arriving at a qualitative understanding, which
is still very incomplete.
By contrast in the ultracold Fermi gases the focus has been
on temperatures associated with the ground state and on arriving at
a more complete quantitative characterization.

This brings us to a major goal of the present paper
which is to suggest new directions in
the cold gas research agenda from the condensed matter
perspective. In particular we wish
to highlight differences and similarities
in the cold gases with the analogous cuprate studies.
A general theme, which takes a cue from the copper oxide
superconductors, is to focus on a characterization of (i) the
fermionic self energy, (ii) the pseudogap phase and (iii)
how superfluid coherence is established and
manifested (in these spectroscopies) at and below $T_c$.

\subsection{Comparing and Contrasting RF With
Photoemission}

\begin{figure*}
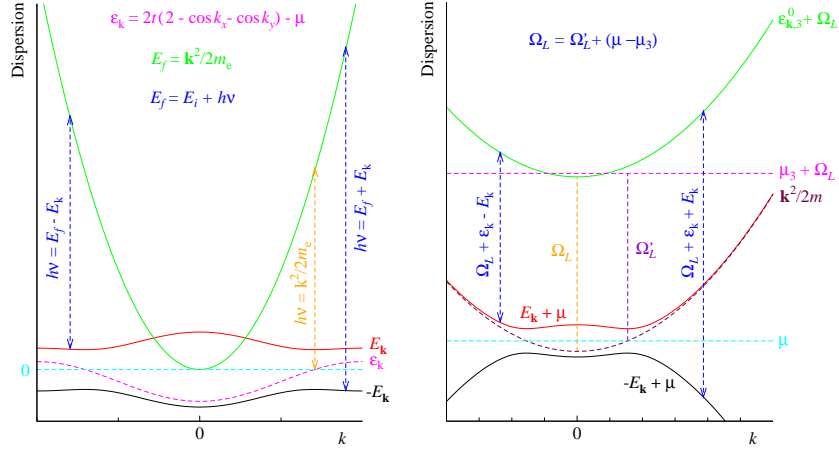

\centerline{\includegraphics[width=2.0in,clip]{ARPES2.eps} \hskip
2ex
\includegraphics[width=2.2in,clip]{RFTransitions.eps}}
%\vskip 1ex
%\centerline{\hspace*{0.in}ARPES\hskip 2.3in RF Transitions}
\caption{
Figure on left:
Energy levels in an ARPES transition. In a paired system there are
two fermionic states which contribute to the photoemitted current. These
correspond to the red and black curves. The upper branch (red curve)
will not be occupied until the temperature is high. Here a tight
binding dispersion $\epsilon_{\bf k}$ is assume for the underlying
non-paired initial state (magenta curve). The cyan dashed line
indicates the Fermi level of the electrons and the green solid line the
dispersion of outgoing electrons.
Figure on right: Energy levels in an RF transition. $\Omega_L$ is the RF
frequency for exciting a free atom from hyperfine level 2 (maroon
line) to level 3 (green line). $\Omega_L^\prime$ is the same energy
but measured relative to the respective chemical potentials. The black
and red curves are the dispersion of the particle and hole branch of a
paired atom in level 2, with energy level given by $\mp \Ek + \mu$,
respectively.
%The RF detuning for excitations from the particle branch
%is
%$\delta\nu = \ek\pm\Ek = \ek_{,3}\pm \Ek -(\mu -\mu_3) =
%\Omega-(\mu -\mu_3)$.
}\label{RFtran}
\end{figure*}

Photoemission and Angle Resolved Photoemission
Spectroscopy (ARPES) have been remarkable tools for characterizing
the cuprate superconductors \cite{arpesanl,arpesstanford_review}.
Here one invokes
the ``sudden'' approximation which corresponds to
the assumption that the electron acquires the photon energy
instantaneously and emerges from the crystal surface immediately. As
a consequence, photoemission is
associated with electrons near the crystal surface. In
addition, only the momentum component in parallel with the surface is
conserved. It follows that ARPES is ideal for layered materials.
The energy levels involved in the ARPES process are shown in
Figure \ref{RFtran}a. Here, and throughout the paper, we define
the quantity $\Ek$ corresponding to the dispersion of the paired
fermions in terms of the usual BCS expression
\begin{equation}
\Ek \equiv \sqrt{ (\ek -\mu)^2 + \Delta^2(T)  }
\label{eq:13}
\end{equation}

Because of the large
photon energy $h\nu$, compared to the electron energy scale inside the
crystal, the final state of the photo-emitted electron is
essentially free so that
the energy conservation constraint is given by
%
%\begin{equation}
$E_i = E_f -h\nu$,
%\end{equation}
%
where
$E_f = \mathbf{k}^2/2m_{\mbox{e}}$
is measured with an energy
analyzer. Here $m_{\mbox{e}}$ denotes electron mass. In turn, the momentum (in the known direction)
has magnitude $k=\sqrt{2m_{\mbox{e}}
E_f}$.
The ARPES spectrum is given by
\begin{equation}
I^{photo} (\mathbf{k},\omega) = M_0(\mathbf{k},\nu) A(\mathbf{k},\omega) f(\omega)
\label{eq:photo1}
\end{equation}
where $M_0(\mathbf{k},\nu)$ is a matrix element which depends on the photon
energy.
%
%The fact that $E_f >> E_i$ and $\mathbf{k}_f = \mathbf{k}_i =
%\mathbf{k}$ is consistent with the fact that the effective electron mass
%inside the crystal is associated with a tight binding model and is much
%heavier than that of free electrons.
%
Apart from the matrix element and the Fermi function $f(\omega)$,
one sees that ARPES
measures the electronic spectral function.
%
%It should be noted that the energy zero in ARPES for both the outgoing
%electrons and the electrons in the crystal is the vacuum energy level at
%the crystal surface, i.e., the chemical potential in the bulk.

The energy levels involved in an RF transition are shown in
Fig.\ref{RFtran}b. Here $\Omega_L$ is the RF frequency for exciting
a free atom from hyperfine level 2 (maroon line) to level 3 (green
line). We neglect final state effects, which will be discussed
later. A significant difference between an RF and ARPES transition
is that in the RF case a dominantly large fraction ($\Omega_L$) of
the photon energy
%$h\nu$
is converted to
excite a fermion from one internal state to another.
As a
consequence, the excited atoms do not have a substantially higher
kinetic energy so that they do not leave the bulk gas immediately after
the transition until they are deliberately released.
%On the other hand,
%the electronic excitation can be regarded as a perfect ``sudden''
%transition (in the absence of final state effects) when compared with
%the relatively slow motion of the atoms.
The energy zero for an RF transition is more conveniently chosen
to be the bottom of the free atom band of state 2. In this
convention, the final state energy is $E_f = \Omega_L +
\epsilon_\mathbf{k}$, where $\ek = k^2/2m$, and the initial state
energy is $E_i = \pm \Ek +\mu$ for the two branches shown in Fig.
1b. Therefore, the same energy conservation constraint emerges
$h\nu = E_f - E_i$. Finally the RF current (which will be derived
in Section \ref{LinearResponse}) is
\begin{equation}
I_0^{RF}(\mathbf{k}, \delta\nu) =\left. \frac{|T_k|^2}{2\pi}
A(\mathbf{k}, \omega) f(\omega)\right|_{\omega=\ek -\delta\nu}
\label{RFc0m}
\end{equation}
where
$|T_k|^2$ is a tunneling matrix element
and, the momentum integrated current, which is the more widely studied form, is
\begin{multline}
I_0^{RF}( \delta\nu)
=\sum_{\bf k}
I_0^{RF}(\mathbf{k}, \delta\nu) \\
= \sum_{\bf k} \left. \frac{|T_k|^2}{2\pi} A(\mathbf{k}, \omega)
f(\omega)\right|_{\omega=\ek -\delta\nu} \label{RFc0}
\end{multline}
We note an important contrast with
Eq.~(\ref{eq:photo1})
because here there is the restriction
$\omega=\ek -\delta\nu$ which (apart from the matrix element effects)
serves to differentiate the photoemission and RF responses.

\subsection{Overview of the Literature on RF Experiments}

Experiments and theory have worked well hand in hand in developing
an understanding of the so-called ``RF pairing gap spectroscopy"
in the atomic Fermi gases. This class of experiments was
originally suggested by Torma and Zoller \cite{Torma} as a method
for establishing the presence of superfluidity. In this context an
equation equivalent to Eq (\ref{RFc0}) was derived. Later work
\cite{JS2,Torma1}, made the observation that these RF experiments,
which reflect the spectral function $A({\bf k}, \omega)$, would 
observe a pairing gap $\Delta(T)$ which may be unrelated to 
superconducting order (except in the strict BCS regime). This was 
the beginning of a recognition that a pseudogap would be present, 
which is associated with stronger-than-BCS attractive 
interactions. Moreover, this pseudogap appears in the ``fermionic 
regime", that is, when the fermionic chemical potential is 
positive \cite{ourreview}.

An experimental ground breaking paper \cite{Grimm4}
reported the first experimental implementation of this pairing gap spectroscopy
in $^6$Li over a range of fields corresponding to the BCS, BEC and unitary
regimes. Accompanying this paper was a theoretical study \cite{Torma2}
by Torma and co-workers based on the BCS-BEC crossover
approach introduced earlier \cite{JS2}, but, importantly,
generalized to include trap effects.
This theoretical scheme is the one that will be the focus of the present paper.
The calculations showed reasonable agreement
with experiment, and subsequent work \cite{heyan} presented more
quantitative comparisons of the spectra along with theoretically-inferred
estimates of the temperature, based on an adiabatic sweep
thermometry \cite{ChenThermo}.  Some of the first evidence that one was, indeed,
observing a pairing gap (or pseudogap) in the normal phase was
presented in Reference \cite{Varenna}, based on this same
thermometric approach and the data of the Innsbruck group \cite{Grimm4}.

In an important contribution Yu and Baym pointed out \cite{YuBaym}
that the theoretical framework described above and summarized in
Eq(\ref{RFc0}) missed what have now become known as ``final state
effects". Moreover, this could be seen most clearly in sum rule
constraints on the RF spectra.  These final state effects can be
understood as follows. Assume as the right panel of Figure~\ref{RFtran} 
that the condensed
phase involves pairing among hyperfine channels 1 and 2 and that
the excited atomic state is associated with hyperfine level 3.
While the attractive interaction $g_{12}$ drives the pairing, the
excited atoms in 3 will also experience a residual interaction
$g_{13}$, which may modify the RF spectra. In this way, these
final state effects yield corrections to the lowest order current,
shown in
Eq (\ref{RFc0}).
Interestingly, the sum rule, now known as the
``clock shift" sum rule \cite{Baym2} shows that the first moment
of the current sums to an internally consistent value, rather than
a pre-determined constant. This will be discussed in Section \ref{SumRule}

A new set of groundbreaking experiments from MIT have introduced a
powerful way of exploiting and enhancing RF spectroscopy first via
tomographic techniques \cite{MITtomo}. With the tomographic scans,
the complication of studying the spectra in a trapped
configuration can now be removed, so that the system is
effectively homogeneous. Also important was the demonstration that
the entire collection of $^6$Li superfluids with hyperfine levels
1 and 2 paired, as well as 1 and 3 as well as 2 and 3, 
are
stable and can be probed in RF spectroscopy with variable RF
transitions, $\Omega_L$ (defined in the right panel of Figure~\ref{RFtran}). In this way one
has, in conjunction with a larger complex of superfluids, a way of
tuning final state effects. Moreover, it was hoped that a proper
choice of the superfluid and the RF transition
can reduce the importance of these final state corrections and
allow one to consider the simpler theory of Eq(\ref{RFc0}).

The theoretical challenge of incorporating final state
contributions has become very topical, in large part because of
the existence of data in effectively ``homogeneous" systems
through these tomographic techniques. It is only in the absence of
a trap that one can readily handle the higher order terms
introduced by Yu and Baym \cite{YuBaym}. With these corrections to
Eq(\ref{RFc0}) one may have a better opportunity to quantitatively
fit the RF spectra. Very nice calculations \cite{Strinati7,Basu}
of $I(\nu)$ in the homogeneous case consider the $T \approx 0 $
superfluid and good agreement with experiment has been
demonstrated \cite{Strinati7}. Subsequent work \cite{ourRF3} has
addressed the entire range of temperatures where one can probe the
RF contributions associated with pre-existing thermally excited
quasi-particles. These are shown as a second branch of RF
transitions in the right panel of Figure~\ref{RFtran}. The body of work
\cite{heyan,heyan2}
at general
temperatures $T$ makes the important point that the presence or
absence of superfluid order (as long as $T < T^*$)
will not lead to fundamentally
different physics. This observation is in contrast to 
alternative calculations \cite{Strinati7,Stoof3,Basu} which
consider only the $T \approx 0 $ superfluid and/or separately the
normal phase.

Along with these new developments has been an experimental
and theoretical focus on
population imbalanced gases
\cite{SR06,SM06,Chien06,Rice2,ChienPRL}.
The observation \cite{KetterleRF} that extreme imbalance
may drive the system to an exotic
normal phase has captured the attention of the community.
This exotic phase appears to be associated \cite{Lobo,Chevy2} with the binding
of a small number of reverse spins to the majority states and this signature
is consistent with
RF experiments, as shown
theoretically \cite{Stoof3,Punk}. It should be stressed that
this binding is not the same as pairing which is a macroscopic
many body phenomenon. But it may, nevertheless, smoothly evolve into
pairing as one varies the concentration of reverse spins
\cite{MITtomoimb}, and in this way diminishes the population
imbalance.

With the growing appreciation for final state effects, an
interesting controversy has recently emerged concerning slightly
different data obtained on the 12 superfluid at unitarity. This
involves the original Innsbruck experiment \cite{Grimm4} and more
recent data from the MIT group \cite{Ketterlepairsize}. The latter
series of studies have led the authors to inquire as to whether
the pairing gap observations reported in Reference
\cite{Grimm4} might instead be associated with final state
effects. We comment on this possibility in Section
\ref{FinalStateCMP} of the paper, where we argue on behalf of the
original interpretation in Reference \cite{Grimm4}.

Finally, recent experiments on $^{40}$K from the JILA group
\cite{Jin6} have now demonstrated that it is possible to
measure the spectral functions directly using momentum resolved RF
pairing gap spectroscopy over a range of magnetic fields
throughout the BCS-BEC crossover. In these recent experiments
\cite{Jin6} the momentum of state 3 atoms is obtained using
time-of-flight imaging, in conjunction with 3D distribution
reconstruction.  Since the 3D gas in a single trap is isotropic,
detailed angular information is irrelevant. There is a substantial
advantage of using $K^{40}$ over the more widely studied $^6$Li
since, for the usual Feshbach resonance around 202 G, there are no
nearby competing resonances to introduce complications from final
state interactions \cite{Strinati7,YuBaym,Punk,Basu, ourRF3}.
This powerful tool, which we have seen has a strong analogy with
ARPES spectra, opens the door for testing the fundamentals of the
many body theory which underly this BCS-BEC crossover. As we show
later in Section \ref{FinalStateCMP} it also helps to remove
ambiguity plaguing the interpretation
\cite{Ketterlepairsize,Grimm4} of momentum integrated RF
experiments by establishing a clear dispersion signature of
pairing.

\subsection{Key Features of ARPES Data on Cuprates}

\begin{figure}
\centerline{\includegraphics[width=2.8in]{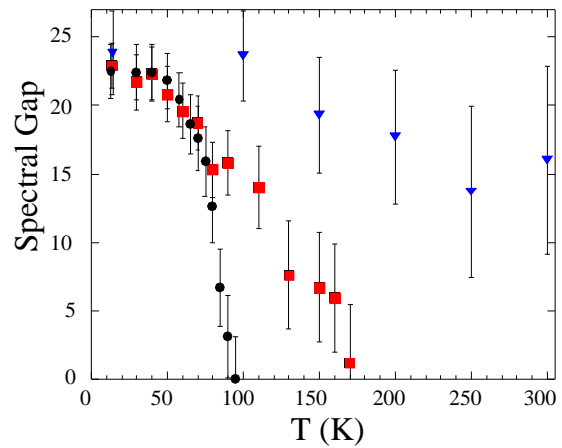}}
\caption{(Color online) Temperature dependence of the excitation gap
from the ARPES measurement for optimally doped (filled black circles, $T_c=87K$),
underdoped (red squares, $T_c=83K$) and highly underdoped (blue inverted
triangles, $T_c=10K$) single-crystal BSCCO samples (taken from
Ref.~ \cite{arpesanl1}). There exists a pseudogap phase above $T_c$ in
the underdoped regime.}
\label{fig:7}
\end{figure}

\begin{figure}
\centerline{\includegraphics[clip,width=2.2in]
{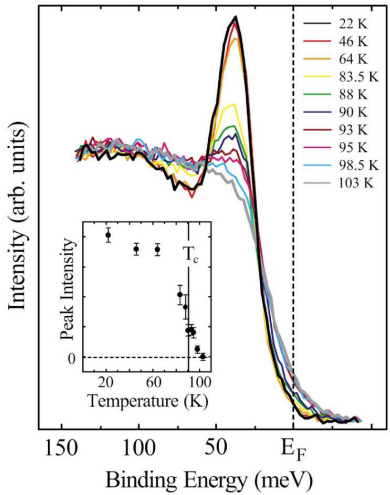}}
\caption{(Color online) Temperature dependent photoemission spectra from
optimally doped Bi2212 ($T_c = 91K$), angle integrated over a
narrow cut at $(\pi, 0$). Inset: superconducting peak intensity vs temperature.
After Ref.\cite{arpesstanford_review}. Note the sharpening of the peaks
as temperature is lowered below $T_c$.}
\label{fig:1}
\end{figure}

We outlined earlier three issues around which much of the
cuprate photoemission studies can be organized. These are characterizations
and modelling of the fermionic self energy, of the pseudogap phase and
of the effects of coherence as the superconductor passes from above to
below $T_c$. Figure \ref{fig:7} is a plot showing the behavior of
the excitation gap which addresses the first of these
three issues. Plotted here is the pairing gap inferred from the leading
edge in the photoemission experiments as a function of temperature.
The temperature
$T^*$ can be read off as the temperature where the gap first appears.
The three different curves correspond to three different doping
concentrations which one can interpret in the framework of BCS-BEC
crossover as corresponding to three different values for the attraction
strength, since they correspond to three different values for
the pairing onset $T^*$.

Several key points can be made. The transition temperatures for phase
coherent order are not evident when one studies the
pairing gap, as shown in the figure except, perhaps, in the sample
with the lowest $T^*$ corresponding to the highest doping concentration.
In this case $T^* \approx T_c$ as in the BCS limit.
In general the higher is $T^*$ the lower is $T_c$ which is, in fact,
consistent with what one expects for BCS-BEC crossover on
a lattice \cite{Chienlattice,Micnaslattice,MicnasRMP}.
Because the behavior varies so smoothly from above to below $T_c$ one
says that the pseudogap smoothly evolves into the excitation gap of
the superfluid phase (sometimes called the
``superconducting gap").  Indeed, independent experiments show that these
two have the same $d$-wave symmetry.

Ideally, one would like to obtain the analogous plots which
show the temperature dependence of the pairing gap in the
cold Fermi gases, using RF spectroscopy in the
BCS, unitary and BEC regimes.
It should be noted that none of the cuprate curves represent the BEC case.
Even though the ratio $T^*/T_c$ can be quite large due to lattice
effects \cite{NSR,Chienlattice}, the high temperature superconductors are all
in the fermionic regime.

In Figure \ref{fig:1} the cuprate photoemission spectra near
optimal doping ($T_c \approx 91K$) are plotted for a range of
different temperatures in order to exhibit the effects of emerging
phase coherence. This figure represents angle integrated spectra
over a cut near the $d$-wave anti-node (where the gap is largest).
What is most striking here is the fact that a sharp quasi-particle
peak emerges only below $T_c$. Above $T_c$ there is a pairing gap,
but it is associated with a relatively poorly defined (gapped)
quasi-particle. In the BCS-BEC crossover scenario, one might view
this as representing short lived, but non-condensed pairs, which
only below $T_c$ can become long lived and stable.

Also of interest in the figure is a feature known as the peak-dip-hump
structure which is associated with superconducting coherence. There is
still some controversy over the origin of these effects, but some
\cite{Pepin}
have
correlated them with specific bosonic modes which couple to
the fermions and appear in
the
self energy.

Again, it would be interesting to have more complete analogous studies using
RF spectra on the cold gases as the system varies from above to
below $T_c$. Just what are the precise signatures of superfluid coherence
and is there evidence that short lived non-condensed pair states become
longer lived below $T_c$, needs to be addressed.
Here one must (perhaps through tomography) overcome the complexity
introduced because these gases are contained in a trap.

\section{General Theoretical Background}

\subsection{BCS Leggett T-matrix Theory}
\label{BCSLeggett}

This paper will address the theory behind RF spectra and
photoemission in the cuprates in the context of one particular
approach to BCS-BEC crossover based on the BCS-Leggett ground
state. Here, however, we generalize to finite temperatures $T$.
There is an alternative approach \cite{Strinati7,Stoof3} based on
the Nozieres Schmitt-Rink scheme \cite{NSR} and which involves
another ground state. Because of the flexibility of the
BCS-Leggett scheme which can readily be generalized to include
trap effects within the superfluid phase,
as well as population imbalance, we choose this alternative.
Another major advantage (from our perspective) is that it is not
plagued by issues associated with a first order transition
\cite{firstordertransitionpapers} at $T_c$. These effects are
related to analogous behavior in mean field theories of the Bose
gas. Interpretation of the cuprate data, which shows a smooth evolution through
$T_c$, would be problematic in the presence of first order effects.

We briefly summarize the key equations which emerge from our $T$ matrix
scheme.
Within the present approach
there there are two contributions to the full $T$-matrix
$t = t_{pg} + t_{sc}$ where $t_{sc}(Q)= -\frac{\Delta_{sc}^2}{T}
\delta(Q)$, where $\Delta_{sc}$ is the superfluid (sc) order parameter.
Similarly, we have two terms for the fermion self energy $\Sigma(K) =
\Sigma_{sc}(K) + \Sigma_{pg} (K) = \sum_Q t(Q)G_{0} (Q-K).$
Here $K$ and $Q$ are four-vectors.
It follows then that
$\Sigma_{sc}(\mb{k},\omega) =
\frac{\Delta_{\mb{k},sc}^2}{\omega +\ek-\mu} $.
Throughout this paper the label $pg$ corresponds to the ``pseudogap" and the
corresponding non condensed pair propagator is given by
%their propagator below $T_c$,
\begin{equation}
t_{pg}(Q)= U/ [1+U \chi(Q)],
\label{eq:14a}
\end{equation}
where the pair susceptibility $\chi(Q)$ has to be properly chosen to
arrive
at the BCS-Leggett ground state
equations. We impose the natural condition that below $T_c$ there is
a vanishing chemical potential for the non-condensed pairs
\begin{equation}
\mu_{pair} = 0
\label{eq:9}
\end{equation}
which
means that $t_{pg}(Q)$ diverges at $Q=0$
when $T\le T_c$. Thus, we approximate \cite{Maly1,Kosztin1} $\Sigma_{pg}(K)$
to yield
\begin{equation}
\Sigma_{pg} (K)\approx -G_{0} (-K) \Delta_{pg}^2 ~~~T \leq T_c\,,
\label{eq:sigma3}
\end{equation}
with
\begin{equation}
\Delta_{pg}^2 \equiv -\sum_{Q\neq 0} t_{pg}(Q).
\label{eq:18}
\end{equation}

It follows that
we have
the usual BCS-like form for the self energy
$\Sigma({\bf k}, \omega )  \approx \Delta^2 / [\omega
+\epsilon_{\bf k}-\mu  ]$ with
$ T \le T_c $
with
\begin{eqnarray}
%\Delta_{mf}(T) &=& \Delta(T)
%\label{eq:new1} \\
\Delta^2(T) &=& \Delta_{pg}^2 (T) + \Delta_{sc}^2(T).
\label{eq:sum}
\end{eqnarray}
As is
consistent with the standard ground state constraints, $\Delta_{pg}$
vanishes at $T \equiv 0 $, where all pairs are condensed.

Using this self energy, one determines $G$ and thereby can evaluate
$t_{pg}$.
Then the condition that the non-condensed pairs have
a gapless excitation spectrum ($\mu_{pair} =0$) becomes
the usual BCS gap equation, except that
it is the excitation gap $\Delta$ and not the order parameter $\Delta_{sc}$
which
appears here.
We then have from
Eq.~(\ref{eq:9})
\begin{equation}
1  + U  \mathop{\sum_{\bf k}}  \frac{1 - 2 f(E_{\bf k})}{2
E_{\bf k}}  = 0,
\qquad  T \le T_c\;. 
\label{eq:gap_equation}
\end{equation}
For consistency we must take for the pair susceptibility
\begin{equation}
\chi(Q)=\sum_{K}G_{0}(Q-K)G(K). 
\label{eq:15}
\end{equation}
Here $G = (G_0^{-1} - \Sigma)^{-1}$ and $G_0$ are the
full and bare Green's functions respectively.

Similarly, using
\begin{equation}
n = 2 \sum_K G(K)
\label{eq:22}
\end{equation}
one derives
\begin{equation}
n  = \sum _{\bf k} \left[ 1 -\frac{\ek - \mu}{\Ek}
+2\frac{\ek - \mu}{\Ek}f(\Ek)  \right]
\label{eq:23}
\end{equation}
which is the natural generalization of
the BCS number equation.
The final set of equations which must be solved is rather simple and given by
Eq.~(\ref{eq:18}),
Eq.~(\ref{eq:gap_equation}), and
Eq.~(\ref{eq:23}).
Note that in the normal state (where $\mu_{pair}$ is nonzero),
Eq.~(\ref{eq:sigma3}) is no longer a good approximation, although
a natural extension can be readily written down \cite{heyan2}.

We stress that the approximation in
Eq.~(\ref{eq:sigma3})
is not central to the physics, but it does greatly simplify the
numerical analysis. One can see that correlations which do not
involve pairing, such as Hartree terms are not included here.
This is what is required to arrive at the BCS-Leggett ground state.
It should be clear that, in principle, the T-matrix approach
discussed here is more general and that
in order to address experiments at a more quantitative level
it will be necessary to go beyond
Eq.~(\ref{eq:sigma3}).
Various groups \cite{Strinati7,Stoof3}
have included these so-called ``$G_0G_0$" contributions
to the pair susceptility $\chi(Q)$ in the T-matrix.
These are particularly important when the pairing is weak
at high temperatures, or for imbalanced gases \cite{Chevy2,Punk,Stoof3},
and, like the Hartree corrections, will ultimately have to be included.

\subsection{Generalization To Include Population Imbalance}

One major advantage of the BCS-Leggett approach is that it
is straightforwardly generalized to include a population imbalanced
superfluid. We begin by summarizing the general equations associated
with the so-called ``Sarma state" which corresponds to a uniformly polarized
BCS superfluid. This is to be distinguished from the 
phase separated state \cite{SR06}.

The gap equation is given by
\begin{eqnarray}
0
%&=&g^{-1}-\sum_{\mathbf{k}}\Big[\frac{f(E_{k\downarrow})}{2E_{k}}-\frac{f(-E_{k\uparrow})}{2E_{k}} \Big] \nonumber \\
%&=&\frac{1}{U}+\sum_{\mathbf{k}}
%\left[\frac{1-f(E_{k\downarrow})-f(E_{k\uparrow})}{2E_{k}}
%\right]\nonumber\\
&=& \frac{1}{U}+\sum_{\mathbf{k}}\frac{1-2\bar{f}(E_{k})}{2E_{k}} \,.
\label{eq:10}
\end{eqnarray}
Here we define the average
\begin{equation}
\bar{f}(x) \equiv
[f(x+h)+f(x-h)]/2,
\label{eq:fbar1}
\end{equation}
where $f(x)$ is the Fermi distribution function.
In addition we define
$\mu=(\mu_{\uparrow}+\mu_{\downarrow})/2$ and
$h=(\mu_{\uparrow} -\mu_{\downarrow})/2$,
%$\xi_{k\uparrow}=-h+\xi_{k}$
%and $\xi_{k\downarrow}=h+\xi _{k}$,
$E_{\bf k} = \sqrt{\xi_{\bf k}^2 +\Delta^2}$, $E_{{\bf k}\uparrow}=-h+E_{\bf k}$ and
$E_{{\bf k}\downarrow}=h+E_{\bf k}$, where $\xi_{\bf k}=\epsilon_{\bf k}-\mu$.

There are now two number equations given by
\begin{subequations}
\label{eq:neq}
\begin{eqnarray}
\label{eq:neqa}
n &=& 2\sum_\mathbf{k} \left[\vk^2 + \frac{\xi_\mathbf{k}}{\Ek}
\bar{f}(\Ek)\right],\\
\delta n &=& \sum_\mathbf{k} [f(\Ek-h)-f(\Ek+h)] \,,
\label{eq:neqb}
\end{eqnarray}
\end{subequations}
where $n= n_\uparrow +n_\downarrow$ is the total atomic density, $\delta
n = n_\uparrow -n_\downarrow >0$ is the number difference and $\delta \equiv \delta
n/n$ is the polarization.
Here the coefficients $u^{2}_{\mathbf{k}}, v^{2}_{\mathbf{k}}=(1\pm
\xi_{\mathbf{k}}/E_{\mathbf{k}})/2$ are formally the same
for both the polarized and unpolarized systems.

Finally, one has to recompute $\Delta_{pg}^2$
\begin{equation}
\Delta_{pg}^2(T)  = \Delta^2 (T) - \Delta_{sc}^2 (T)
= -\sum_{Q\neq 0} t_{pg}(Q)
\label{eq:Dpg}
\end{equation}

This quantity can
be obtained directly from an equation in the same
form as
Eq.~(\ref{eq:14a})
except that the pair
susceptibility appearing here satisfies
\begin{equation}
\chi(Q)=\frac{1}{2}\big[\chi_{\uparrow\downarrow}(Q)+
\chi_{\downarrow\uparrow}(Q)\big]
\label{eq:2h}
\end{equation}
where as before we have the product of one dressed and one bare Green's function
\begin{subequations}
\label{eq:3h}
\begin{eqnarray}
\label{eq:3ah}
\chi_{\uparrow\downarrow}(Q)&=&\sum_{K}G_{0\uparrow}(Q-K)G_{\downarrow}(K)  \\
\chi_{\downarrow\uparrow}(Q)&=&\sum_{K}G_{0\downarrow}(Q-K)G_{\uparrow}(K)
\label{eq:3bh}
\end{eqnarray}
\end{subequations}
Further details are presented in Reference \cite{heyan2}.

\subsection{Linear Response Theory and RF}
\label{LinearResponse}

In the RF experiments \cite{Grimm4a}, one focuses on three different
atomic hyperfine states of the $^6$Li atom.  The two lowest states,
$\ket1$ and $\ket2$, participate in the superfluid pairing.
These correspond to $\uparrow$ and $\downarrow$.
%, and are described by the Hamiltonian $H-\mu N$.
The higher state, $\ket3$, is
effectively a free atom excitation level; it is unoccupied initially.
An RF laser field, at sufficiently large frequency, will drive atoms
from state $\ket2$ to $\ket3$.

We begin with the usual grand canonical Hamiltonian $H-\mu N$
which describes states $\ket1$ and $\ket2$. We have described in
Section \ref{BCSLeggett} the procedure for handling pairing
correlations in this 12 channel. The Hamiltonian describing state
$\ket3$ is given by
$$H_3-\mu_3 N_3=
\sum_{\mathbf{k}} (\epsilon_{\mathbf{k}}-\mu_3)
c_{3,\mathbf{k}}^{\dagger} c^{}_{3,\mathbf{k}}$$ where
$\epsilon_{\bf k}$ is the atomic kinetic energy,
$c^{}_{3,\mathbf{k}}$ is the annihilation operator for state
$\ket3$,
%$\Omega_{L}$ is the energy splitting between $\ket3$ and
% $\ket{2}$, 
and $\mu_3$ is the chemical potential of $\ket3$.

In addition, there is a transfer matrix element $T_{\bf k,p}$ from
$\ket2$ to $\ket3$ given by
$$  H_T  =\sum_{\bf k,p}(T_{\bf k,p}\,c_{3,\bf p}^\dag c_{2,\bf k}^{}
+h.c.) $$
For plane wave states,
$T_{\bf k,p} = T_{23}\delta({\bf q}_L+{\bf k}-{\bf
p})\delta(\omega_{\bf {kp}} - \Omega_L)$.
Here $q_L \approx 0$ and $\Omega_L$ are the momentum and energy of
the RF laser field, and $\omega_{\bf {kp}}$ is the energy
difference between the initial and final states.
In what follows we will set the magnitude of the tunneling matrix
to unity, without loss of generality.
It should be
stressed that unlike conventional SN tunneling, here one requires
not only conservation of energy but also conservation of momentum.

The RF current is defined as
$$I=-\langle \dot{N}_2\rangle=-i\langle
[H,N_2]\rangle.$$ Using standard linear response theory one finds
$$I(\nu)=-\frac{1}{\pi}{\mbox{Im}}[D^{R}(\Omega_L+\mu-\mu_3)].$$
Here we introduce the
retarded response function $D^{R}(\omega) \equiv D(i\omega_n
\rightarrow \omega+i0^+)$.

At the lowest order of approximation the linear response kernel
$D$ can be expressed in terms of single particle Green's functions
as $$D_0(i\omega_n)= T \sum_{K}G_0^{(2)}(K)G^{(3)}(K+Q),$$ where
$K=(\mathbf{k},\omega_n)$ and $Q=(\mathbf{0},\Omega_n)$. (We use
the convention $\hbar=k_B=1$). The Green's function can then be expressed
in terms of spectral functions. After Matsubara summation we
obtain
%the result can be written in
%terms of spectral functions as

\begin{equation}
I_0(\nu) =\frac{1}{4\pi^2}\int \!\mathrm{d}\epsilon\sum_{{\bf k}} A({\bf
k},\epsilon)A_3({\bf k},\bar{\epsilon})
\left[f(\bar{\epsilon})-f(\epsilon)\right] \,,
\end{equation}
with $\bar{\epsilon}=\epsilon+\nu+\mu-\mu_3$, $\nu$ is the RF detuning
 and $f(x)$ is the Fermi
distribution function.

The spectral function for state $\ket3$ is $A_3({\bf
k},\epsilon)=2\pi \delta(\epsilon-\xikk)$. We see from the above
equation that these RF experiments depend on
$A({\bf k},\epsilon) \equiv -2\, \mathrm{Im}\, G({\bf
k},\epsilon+i 0^+)$ which is the spectral function associated with
the superfluid component $\ket2$. Then the lowest order RF current can be written as
\begin{equation}
I_0(\nu) =-\frac{1}{2\pi}\sum_{{\bf k}}A({\bf
k},\xik-\nu)\left[f(\xik-\nu)-f(\xikk)\right] \,,
\end{equation}
In practice, state $3$ is unoccupied, thus the second Fermi
function in brackets vanishes. In this way we have for the
momentum-resolved current

\begin{eqnarray}
I_0(\mathbf{k},\nu)=-\frac{1}{2\pi}A({\bf k},\xik-\nu)f(\xik-\nu).
\label{eq:RFcurrent}
\end{eqnarray}

\subsection{BCS-Leggett Model for Self Energy}

The current $I_0(\nu)$ at the leading order level depends on
the fermionic spectral function, which, in turn depends on
the fermionic self energy.  In this section we discuss
the nature of the self energy which will enter into an analysis
of both photoemission and RF spectroscopy.

To arrive at the BCS-Leggett ground state equations, we have seen that the self energy
is given by
$\Sigma(\mb{k},\omega) =
\Sigma_{sc}(\mb{k},\omega) +
\Sigma_{pg}(\mb{k},\omega)$ 
where
\begin{eqnarray}
\Sigma(\mb{k},\omega) &=&
\frac{\Delta_{\mb{k},sc}^2}{\omega +\ek-\mu}
%\Sigma_{sc}(\mb{k},\omega) +
+\Sigma_{pg}(\mb{k},\omega) \label{selfE3} \\
&=&
\frac{\Delta_{\mb{k},sc}^2}{\omega +\ek-\mu} +
\frac{\Delta_{\mb{k},pg}^2}{\omega +\ek-\mu}
\mbox{ },\mbox{ }T \leq T_c \\
\label{selfE2} &=& \frac{\Delta_{\mb{k},pg}^2}{\omega +\ek-\mu}
\mbox{ },\mbox{ }T
> T_c  \label{selfE1}
\end{eqnarray}

These equations follow, provided
one makes the approximation contained in Eq.~(\ref{eq:sigma3}). In invoking
this approximation we are in effect ignoring the difference between
condensed and non-condensed pairs which cannot be strictly correct.
The simplest
correction
to $\Sigma_{pg}$ (which should apply above and below $T_c$) is to write
\begin{equation}
\Sigma_{pg}(\mb{k},\omega) \approx
\frac{\Delta_{\mb{k},pg}^2}{\omega+\ek-\mu+i\gamma} +\Sigma_0 (\mb{k},\omega) \,.
\label{SigmaPG_Model_Eq}
\end{equation}
Here the broadening $\gamma \ne 0$ and ``incoherent'' background
contribution $\Sigma_0$ reflect the fact that noncondensed pairs do not lead to
\textit{true} off-diagonal long-range order.
While we can think of $\gamma$ as a phenomenological parameter in the
spirit of the high $T_c$ literature \cite{Normanarcs,Chubukov2}
there is a microscopic basis for considering this broadened BCS
form \cite{Maly1,Maly2}.
The precise value of $\gamma$, and even its $T$-dependence is not
particularly important for the present purposes, as long as it is non-zero at finite $T$.
For simplicity we will take $\gamma$ as a temperature independent
constant throughout this paper.
By contrast $\Sigma_{sc}$ is associated with long-lived
condensed Cooper pairs, and
is
similar to $\Sigma_{pg}$ but without the broadening.

It is important to stress that this same self energy model has been
applied to describe the spectral function in the pseudogap
\cite{Chen4,Normanarcs,Chubukov2} and the superfluid phases
\cite{FermiArcs} of the high temperature superconductors, where
here $\Sigma_0(\mb{k},\omega) \equiv - i \Gamma_0$ is taken to be an imaginary
constant. In the cuprate literature, it has been argued that
Eq.~(\ref{SigmaPG_Model_Eq}) is appropriate to the normal
phase and the onset of coherence coincides with a
dramatic decrease in $\gamma$ below $T_c$. This leads to
a subtle and important controversy about which we wish to comment
more in the context of Section \ref{PhotoEXP}. Our own perspective
is that Eq.~(\ref{selfE3}) in conjunction with
Eq.~(\ref{SigmaPG_Model_Eq})
is the appropriate
starting point. That is, there are two contributions to
the self energy below $T_c$ and only one above.
Thus one should not argue
that $\gamma$ precisely vanishes at $T_c$ but rather there is a continuous
conversion from non-condensed to condensed pairs as $T$ is lowered
within the superfluid phase. The
non-condensed pairs below $T_c$ have finite lifetime
while the condensed pairs do not.

The resulting spectral function, based on
Eq.~(\ref{SigmaPG_Model_Eq})
and
Eq.~(\ref{selfE3}) 
is
given by
\be A({\bf k},\epsilon)=\frac{2\Delta_{pg}^{2}\gamma
(\epsilon+\xi_{\bf k})^2}{(\epsilon+\xi_{\bf
 k})^2(\epsilon^2-E_{\bf k}^{2})^2+\gamma^2(\epsilon^2-\xi_{\bf
 k}^2-\Delta_{sc}^{2})^2} \,.
 \label{spec}
\ee
Here, for convenience we do not show the effects of
the $\Sigma_0$ term.
%Here $\xi_{\bf k} = \epsilon_{\bf k}-\mu$.  $E_{\bf k} = \sqrt{
%\xi_{\bf k}^2 + \Delta^2 (T) }$ is the quasiparticle
%dispersion, where $\Delta^2(T) = \Delta_{sc}^2(T) + \Delta_{pg}^2(T)$.
Above $T_c$, Eq.~(\ref{spec}) can
be used with $\Delta_{sc} = 0$.
It can be seen that this spectral
function at all ${\bf k}$ contains a zero at $\epsilon =-\xi_{\bf k}$ below $T_c$,
whereas it has no zero above $T_c$.
This means that a clear signature of phase coherence is
present, as long as $\gamma \neq 0$.
%Because the energy level difference $\omega_{23}$ ($ \approx 80 $ MHz)
%is so large compared to other energy scales in the problem, the state
%$\ket3$ is initially empty. It is reasonable to set $f(\epsilon_{\bf
%k}+\omega_{23}- \mu_3) = 0$ in Eq.~(\ref{eq:4}).

This analysis may be readily generalized to include the effects of
population imbalance.
%
%We have $\Sigma = \Sigma_{pg} +
%\Sigma_{sc}$, where $\Sigma_{pg} ({\bf k}, \epsilon) =
%\frac{\Delta_{pg}^2}{\epsilon +\xi_{k,1} +i\gamma} $ and
%$\Sigma_{sc}({\bf k}, \epsilon) =
%\frac{\Delta_{sc}^2}{\epsilon+\xi_{k,1}}$.
We have
for the spectral function of the minority
\be A_2({\bf k},\epsilon)=\frac{2\Delta_{pg}^{2}\gamma
(\epsilon'+\xi_{\bf k})^2}{(\epsilon'+\xi_{\bf
 k})^2(\epsilon'^2-E_{\bf k}^{2})^2+\gamma^2(\epsilon'^2-\xi_{\bf
 k}^2-\Delta_{sc}^{2})^2} \,.
\ee
with $\epsilon'=\epsilon-h$.

Similarly, the spectral function for the majority is
\be A_1({\bf k},\epsilon)=\frac{2\Delta_{pg}^{2}\gamma
(\epsilon''+\xi_{\bf k})^2}{(\epsilon''+\xi_{\bf
 k})^2(\epsilon''^2-E_{\bf k}^{2})^2+\gamma^2(\epsilon''^2-\xi_{\bf
 k}^2-\Delta_{sc}^{2})^2} \,.
\ee
with $\epsilon''=\epsilon+h$.

There are instances where it is problematic to include the
effects associated with the finite lifetime $\gamma$. This occurs
when we compute the effects of
final state interactions.
At this strict ``mean field " (mf) level
we
drop the factor $\gamma$, thereby, losing the
distinction between condensed and
non-condensed pairs.
In this case the spectral function (which we display here for
the polarized gas case) is associated with majority (1) and minority (2)
contributions: 
%
%\begin{equation}
%G^{mf}(K)
% =  \left(
%  \frac{\uksq}{i\omega-\Ek} + \frac{\vksq}{ i\omega+\Ek} \right)
%\end{equation}
%
$$A_1^{mf}(\mathbf{k},\omega) = 2\pi[\uksq \delta(\omega-\Eku)
+ \vksq \delta(\omega+\Ekd) ],$$
$$A_2^{mf}(\mathbf{k},\omega) = 2\pi[\uksq \delta(\omega-\Ekd)
+ \vksq \delta(\omega+\Eku) ],$$ with $\Eku=\Ek-h$ and
$\Ekd=\Ek+h$.
We stress again, however, that here $\Delta$ contains non-condensed
pair effects through
Eq.~(\ref{eq:sum}).

\subsection{Analytical Results for the Leading Order RF Current}
\label{sec:analy}

It is possible to obtain analytical results for the leading
order current at general temperatures $T$ in
this strict mean field theory. Here
one can integrate over the momentum to find
 \ba
 I_0(\nu)&=&\frac{1}{8\pi^2}\frac{\Delta^2}{\nu^2}[1-f(E_0)]k_0\,,\quad (\nu>\nu_1)\\
I_0(\nu)&=&\frac{1}{8\pi^2}\frac{\Delta^2}{\nu^2}f(E_0)k_0\,,\quad
(\nu_2<\nu<0) \ea with
$$
E_0=\bigg|\frac{\nu^2+\Delta^2}{2\nu}\bigg|\,, \quad
k_0^2=\mu+\frac{\nu^2-\Delta^2}{2\nu}.$$
The frequency regimes associated with the
negative and positive continua are 
given by $-(\sqrt{\mu^2 + \Delta^2} +\mu) \le \nu \le 0$
and $\nu \ge \sqrt{\mu^2
+ \Delta^2} -\mu $.
In the above equations 
$ \nu_2 \equiv -(\sqrt{\mu^2 + \Delta^2} +\mu)$ and
$ \nu_1 \equiv \sqrt{\mu^2 + \Delta^2} -\mu$. 
It can be seen that there are contributions for both negative and positive
detuning.
At strictly zero temperature, the Fermi function vanishes,
and we have only the positive
continuum
\ba
 I_0(\nu)&=&\frac{1}{8\pi^2}\frac{\Delta^2}{\nu^2}\sqrt{\mu+\frac{\nu^2-\Delta^2}{2\nu}}. 
 \ea
It is possible to write down a closed form expression for
the polarized case at general $T$ as well. We do not show
it here because the expressions are more cumbersome.
We defer this to Appendix \ref{App:analy}.

\subsection{ Behavior in Traps In Absence of Final State Effects}

\begin{figure*}
\centerline{\includegraphics[width=5.5in,clip]{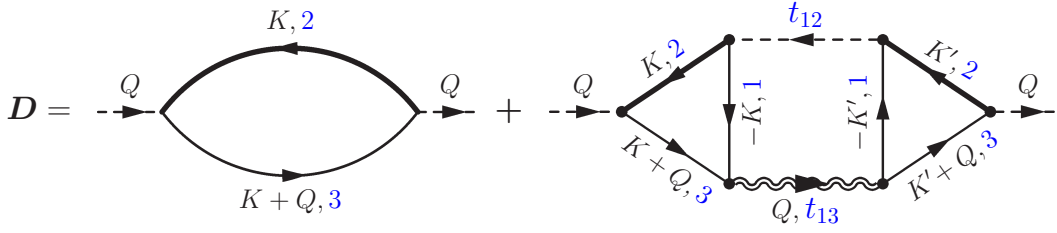}}
\caption{Feynman diagrams for the RF response function
$D(Q)$. The left bubble is the lowest order $D_0$, whereas the right
diagram, $D_{AL}$, is associated with final state effects. Here thin
(thick) lines stand for bare (full) fermion propagators, the dashed
line for $t_{12}$, approximated as the condensate, and double wiggly
line for $t_{13}$. The numbers in blue indicate the hyperfine
levels. Here $Q=(i\Omega_n, \mathbf{0})$ for the RF field.}
\label{fig:Diagrams}
\end{figure*}

Once the trap is
incorporated, one has to solve for the current at each position
$r$ and then integrate in the form
\begin{equation}
I_{\sigma}(\nu) = \int \mathrm{d}^3r\, I(r,
\nu)n_{\sigma}(r)
\end{equation}
where here
$n_{\sigma}(r)$
represents the particle density within the trap and
$\sigma=1,2$ are the different hyperfine levels of the
superfluid.
To handle trap effects we assume a spherically symmetrical harmonic
oscillator potential $V(r) = m\bar{\omega}^2 r^2/2$.  The density,
excitation gap and chemical potential which vary along the radius
can be determined \cite{heyan2} using the local density approximation
(LDA).

It should be stressed that the density and gap profiles
($n_{\sigma}(r)$ and $\Delta(r)$) in general involve
pseudogap or non-condensed pair effects. The strict mean
field theory, which often gives a reasonable approximation
to the spectral functions, is not adequate for obtaining
these trap profiles. Thus even when analyzing tomographic
RF data one has to include the full effects of these pair
excitations \cite{heyan2}, effectively through
$\Delta_{pg}^2(r)$ and non-zero $\mu_{pair}(r)$.

\section{ Final State Effects in Homogeneous Unpolarized System}
\label{FinalStateTheory}

We now turn to the inclusion of final state effects which go
beyond the leading order diagram. It is
complicated to handle these contributions for the inhomogeneous case.
Thus we focus here on treating the RF current as if the system
were homogeneous.
We formulate the finite $T$, RF problem
using a diagrammatic scheme.
The full diagram set for the RF response function, $D(Q)$, is shown
in Fig.~1.
The leading order term $D_0$ appears as the first term on
the right hand size and the second contribution is
associated with
the Aslamazov-Larkin (AL) diagram (called $D_{AL}$).
The full RF
current, given by the retarded response function, is $I(\nu) \equiv
-(1/\pi)\,\mbox{Im}\,D^R(\Omega)$, where $\Omega\equiv\nu+\mu-\mu_3$.

The approximation compatible with
Eq.~(\ref{eq:sigma3})
is effectively
equivalent to treating the
$D_{AL}$ in Fig.~1 at the BCS
mean-field level, leading to the opposite momenta $\pm K$ for particles
1 and 2 in the diagram.
$D_{AL}(Q)$, depends
on $\Delta$, not $\Delta_{sc}$, and incorporates final-state effects via
the interactions $g_{12}$ between 1 and 2 and $g_{13}$ between 1 and 3.
We neglect the effects arising from the interaction between 2 and 3.
This is consistent with the approach in Ref.~\cite{Baym2}. This
second term has appeared previously in studies of the superfluid density
\cite{Chen2}.
Our formulation of the finite $T$, RF problem
can be made compatible with the
diagrams in Ref.~\cite{Strinati7}, although attention in that
paper was restricted to very low temperatures.  
Our diagrammatic scheme reduces at $T=0$ to the approach of
Ref.~\cite{Basu}.

In order to evaluate the AL term, we begin by writing out
the relevant T-matrices
\begin{eqnarray}
t^{-1}_{12}(Q)&=&g_{12}^{-1}+\sum_K G_1(K)G_2^0(Q-K)\\
t^{-1}_{13}(Q)&=&g_{13}^{-1}+\sum_K G_1(K)G_3^0(Q-K)
\end{eqnarray}
where
$g_{12}$  and $g_{13}$ parameterize the interaction between 1 and 2
and 1 and 3, respectively.
We can also introduce the $s$-wave scattering
lengths, $a_{13}$ (and
$a_{12}$) in the 1-3 (and 1-2)
channels, respectively.

Thus
\begin{equation}
t^{-1}_{13}(Q)
%&&=&\frac{1}{g_{13}^{}} + \sum_K G^{(2)}(K)G_0^{(3)}(Q-K)\nonumber\\
=\frac{m}{4\pi a_{13}} + \chi_{13}(Q)
\end{equation}
where
\ba \chi_{13}^{}(Q)&=&-\sumk
\Big[\frac{1-f(\Ek)-f(\xikk)}{i\Omega_n-\Ek-\xikk}\uksq\nonumber\\
&+&\frac{f(\Ek)-f(\xikk)}{i\Omega_n+\Ek-\xikk}\vksq
 +\frac{1}{2\epsilon_\mathbf{k}^0} \Big]
 \ea

The AL diagram yields
\begin{eqnarray}
D_{AL}(Q)=\Big[\sum_K F(K)G_3^0(K+Q)\Big]^2 t_{13}(Q)\,.
\label{eq:D_AL}
\end{eqnarray}

where
\begin{equation}
F(K)\equiv-\Delta G^{(2)}(K) G_0^{(1)}(-K) = \frac{\Delta} {(i\omega_l)^2-\Ek^2}
\end{equation}

This contribution can be rewritten
\ba
D_{AL}(Q)
%&=&\left[ -\sum_K F(K)  G_0^{(3)}(K+Q)  \right]^2 t_{13}(Q)\nonumber\\
&\equiv&D_2^2 (Q)t_{13}(Q) \,, \ea where we have defined

$D_2(Q) \equiv \sum_K F(K)G_3^0(K+Q)$
\begin{equation}
  =\sum_K\frac{\Delta}{2E_k}\Big[\frac{1\!-\!f(E_k)\!-\!f(\xi_{k,3})}
  {i\Omega_n-E_k-\xi_{k,3}}
  -\frac{f(E_k)-f(\xi_{k,3})}{i\Omega_n\!+E_k-\xi_{k,3}}\Big]
\end{equation}

%

%We also define
%$D_1(Q) \equiv \sum_K G_1(K)G_3^0(Q-K)-\sum_\mathbf{k} (1/2\epsilon_k) =$
%
%\begin{equation}
%\sum_K\Big[\frac{f(E_k)\!+\!f(\xi_{k,3})\!-\!1}{i\Omega_n-E_k-\xi_{k,3}}u_k^2
%+\frac{f(\xi_{k,3})-f(E_k)}{i\Omega_n\!+\!E_k\!-\!\xi_{k,3}}v_k^2\Big]
%-\sum_\mathbf{k} \frac{m}{k^2}\,.
%\label{eq:D1}
%\end{equation}
%
Then the full set of diagrams shown in Figure \ref{fig:Diagrams}
can be combined to yield
\begin{equation}
D(Q )  =D_0 (Q) +  \frac{[D_2(Q)]^2} {m/4\pi a_{13} +
\chi_{1,3}(Q)}\,, \label{eq:D}
\end{equation}

After analytical continuation and change of variables, we have $\Omega
\pm \Ek -\xi_{k,3} = \nu \pm \Ek -\xi_k$. Importantly, the
denominators here are the same as those which appear in
$t_{12}$. Furthermore, at $\nu=0$, $f(\xi_{k,3})$ is cancelled out so that
\begin{equation}
  t^{-1}_{13}(0) =
  (g_{13}^{-1} - g_{12}^{-1}) + t_{12}^{-1}(0) = g_{13}^{-1} - g_{12}^{-1}.
\end{equation}
It follows that the complex functions $D_0(Q)$, $\chi_{1,3}(Q)$,
and $D_2(Q)$ are the same as their wave function calculation
counterparts \cite{Basu} when the pairing gap $\Delta$ is chosen
to be order parameter $\Delta_{sc}$ and $T=0$. It is $\nu$ not
$\Omega$ that should be identified with the experimental RF
detuning.

After some straightforward algebra (with details in Appendix A), we
find for the RF current
%$I(\nu)=(n_2-n_3)\delta(\nu)$. In general case, we have
\begin{eqnarray}
\lefteqn{I(\nu) = \left[\frac{1}{g_{12}} - \frac{1}{g_{13}}\right]^2\!\!
\frac{I_0(\nu)}{|t^{-1,R}_{13}(\nu)|^2}}\nonumber\\
&=& -\frac{1}{\pi}\left(\frac{m}{4\pi a_{13}} - \frac{m}{4\pi
a_{12}}\right)^2
    \frac{\mbox{Im} D_0^R(\nu)}{|t_{13}^{-1,R}(\nu)|^2}\nonumber\\
&&= -\frac{1}{\pi} \left[\frac{m}{4\pi a_{13}} - \frac{m}{4\pi
a_{12}}\right]^2 \!\frac{\Delta^2}{\nu^2}\, \mbox{Im}\,
t^R_{13}(\nu), \label{eq:I}
\end{eqnarray}
Moreover, in the special case, when $a_{13} = a_{12}$ then
$I(\nu)=(n_2-n_3)\delta(\nu)$.
Equations (\ref{eq:I}) are the central result. It should be
clear that final state effects in the RF current directly reflect the
$T$-matrix in the 1-3 channel.  In general, features in the RF spectra
derive from the poles and imaginary parts of
$D_0(Q)$, $\chi_{1,3}(Q)$,
and $D_2(Q)$.
%Eqs.~(\ref{eq:D})-(\ref{eq:D1}).

The spectrum may contain a bound state
associated with poles at $\nu_0$ in $t_{13}$, as determined by $
t^{-1}_{13}(\nu_0) = 0$.  This leads to the so called ``bound-bound''
transition. In addition, there is a continuum associated with both the
numerator and denominator in the first of Eqs.~(\ref{eq:I}), with each
contribution spanned by the limits of $\nu = \xi_k\pm E_k$, i.e.,
$-(\sqrt{\mu^2 + \Delta^2} +\mu) \le \nu \le 0$ and $\nu \ge \sqrt{\mu^2
  + \Delta^2} -\mu $.  The continuum at positive frequencies is
primarily associated with breaking a pair and promoting the state 2 to
state 3. This represents the so-called ``bound-free'' transition. On the
negative detuning side, the continuum is primarily associated with
promoting to state 3 an already existing thermally excited 2
particle. The spectral weight of the negative continuum vanishes
exponentially at low $T$.

\subsection{Sum Rules}
\label{SumRule}

Of importance, in assessing any theoretical framework for computing the RF
current are the two sum rules associated with the total integrated
current and the first moment or ``clock shift'' \cite{Baym2}. Using the
Kramers-Kronig relations between $\mbox{Re}\,t^R_{13}$ and
$\mbox{Im}\,t^R_{13}$, we prove in Appendix \ref{App:C}
that, not only in the ground
state, but also at finite temperature, Eq.~(\ref{eq:I}) satisfies
\begin{eqnarray}
\int d\nu  \, I(\nu) &=& n_2 - n_3 \,,\\
\int\! d\nu \,\nu\, I(\nu)
%&=& \Delta^2 (g^{-1}_{12}-g^{-1}_{13}) \nonumber \\
&=& \Delta^2\frac{m}{4\pi} \left(\frac{1}{a_{12}} -
\frac{1}{a_{13}}\right)
\end{eqnarray}
where $n_2$ and $n_3 (=0)$ are the density of state 2 and 3 atoms,
respectively.  In this way we find for the clock shift a result
which we write (for general polarizations, associated with
the subscript $\sigma$) in the form:
\begin{equation}
\bar{\nu_{\sigma}} = \frac{\int d\nu \,\nu I_{\sigma}(\nu)} {\int
d\nu \,I_{\sigma}(\nu)}
%&=&  \frac{ \Delta^2}{n_{\sigma} (g^{-1}_{12}-g^{-1}_{13}) \nonumber\\
=\frac{ \Delta^2}{n_{\sigma} - n_3} \frac{m}{4\pi}
\left(\frac{1}{a_{12}} - \frac{1}{a_{13}}\right)\,, \label{eq:14}
\end{equation}
In the unpolarized case,
this agrees with Ref.~\cite{Baym2}.
This sum rule is satisfied only when
$a_{13}\neq 0$ and when both diagrammatic contributions are included. It is
easy to show that at large $\nu$, $I_0(\nu) \sim \nu^{-3/2}$,
$\mbox{Im}\,t_{13}^R \sim \nu^{-1/2}$, so that $I(\nu) \sim \nu^{-5/2}$,
in agreement with Ref.~\cite{Strinati7}.  Clearly, the first
moment of $I(\nu)$ is integrable, whereas the first moment of $I_0(\nu)$
is not.  Finally, Eq.~(\ref{eq:I}) reveals that the spectral weight
(including possible bound states) away from $\nu =0$ will disappear when
the gap $\Delta$ vanishes.

\section{Physical Picture and Implications}

\begin{figure}
\includegraphics[width=3.3in,clip]
{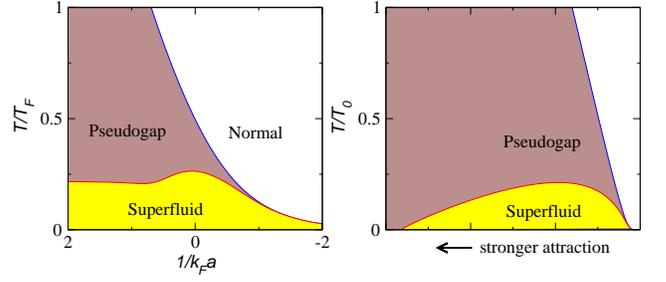}
\caption{Phase diagram showing $T_c$ and $T^*$ for homogeneous $s$-wave
Fermi gas superfluid (left)
and for $d$ wave superfluid on a quasi two dimensional lattice. From
Reference \cite{Chienlattice2}. Note that the BEC asymptote is
finite in a Fermi gas and zero in the lattice case. Because the lattice phase
diagram shows similarity to that of the cuprates,
in future experiments it will be
important to study the $d$-wave generalization of the attractive
Hubbard model on an optical lattice.}
\label{fig:40a}
\end{figure}

\begin{figure}
\includegraphics[width=3.3in,clip]
{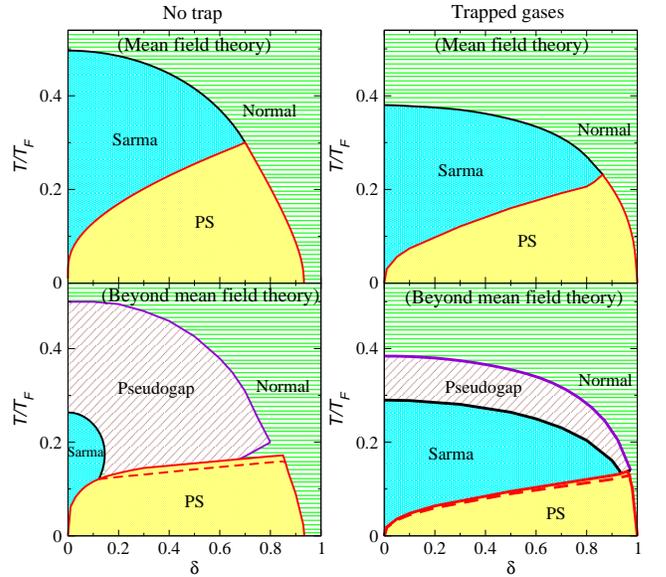}
%{Summary_cmp_imb.eps}
\caption{This summarizes the phase diagrams for polarized gases with and without
a trap and with and without pairing fluctuations. The figure is based on
Refs.~\cite{ChienPRL} and \cite{heyan2}. The figures on the left
are for homogeneous case and on the right for trapped case. The mean field figures
at the top show the reported tri-critical point. However, fluctuations (in
the lower two plots) depress the superfluid phases.}
\label{fig:40b}
\end{figure}

In this section we lay the groundwork for a comparison between theory and
experiment, which is presented in the following
section. We address the various phase diagrams for the 
population balanced Fermi gases, including
the ($d$-wave) lattice case, as well as for the imbalanced systems.
We analyze a pedagogically useful set of figures which 
lay out the general behavior of the RF spectra with and without final
state effects and with and without a trap. Importantly we compare photoemission-based
plots for the same parameter set as RF-based plots and address the key 
signatures of emerging superfluid coherence as one goes from above to below
$T_c$.

\subsection{Phase diagrams}
The relevant phase diagrams to be used 
and referred to in our RF calculations have
been obtained elsewhere. Shown here are the curves for
$T_c$ and $T^*$. The latter represents the boundary
between the pseudogap and normal phases. Figure \ref{fig:40a} compares the phase diagram
for an $s$-wave paired Fermi gas (left) and for a $d$-wave paired fermion
system (right) on a quasi two dimensional lattice. The $s$-wave
gas case is closely analogous to results obtained using the approach of
Refs.~ \cite{NSR} and 
\cite{Strinati4}. 
The $d$-wave case was discussed earlier
in Ref.~\cite{Chen1} and more recently in the context of optical
lattice calculations in Refs.~\cite{Chienlattice} and
\cite{Chienlattice2}.
The seminal Nozieres Schmitt-Rink paper pointed out a key fact which identifies 
a notable difference between the lattice and gas cases: the BEC limit
has an asymptote of $T_c \rightarrow 0$ in the case of a lattice, whereas
it is finite in a gas. Thus
there is a relatively larger separation between $T^*$ and $T_c$ when fermions
are present on a lattice as shown in the figure. 
We note that the $d$-wave case has a number of features in common with the
counterpart phase diagram of the cuprates \cite{Chen1,Chienlattice,Chienlattice2}.

The phase diagrams for polarized ($s$-wave) Fermi gases are shown in
Figure \ref{fig:40b}. The four panels correspond to the effects of including
(or not) a trap and to the effects of including (or not) pairing fluctuations
beyond strict mean field theory, which enter in the theory through the
parameter $\Delta_{pg}^2$. When we discuss the RF behavior of polarized
gases we will use the full beyond-mean-field theory phase diagrams, although
some of the calculations of the spectral function are performed at the strict
mean field level.  

Beyond the normal phase, there are three phases which
appear \cite{ChienPRL,heyan2}: the Sarma phase, a phase separated (PS) state 
and a pseudogapped normal state, as indicated.  We note that the treatment
of the normal component of the
phase separated state does not include correlations beyond those 
accounted for by $\Delta$. As a result, these calculations 
overestimate the range of stability of phase separation.
This issue has been nicely discussed in the theoretical literature 
\cite{Lobo,Chevy2} with implications for RF spectra
as well \cite{Punk,Stoof3}.
The Sarma phase should be considered as the more correctly treated here and one notes
an important finding: that in the absence of a trap the regime of
stability of the Sarma state is greated reduced.
This more restricted stability (seen by comparing the two lower figures) is
associated with the fact that the excess majority fermions can be
accomodated more readily in different spatial regions in a trap.
The maximum polarization of this homogeneous Sarma phase
is around $\delta = 0.2$ which is close
to that reported  
experimentally \cite{MITphase}.

\subsection{Comparison with the Cuprates}

\begin{figure}
\centerline{\includegraphics[clip,width=2.4in]
{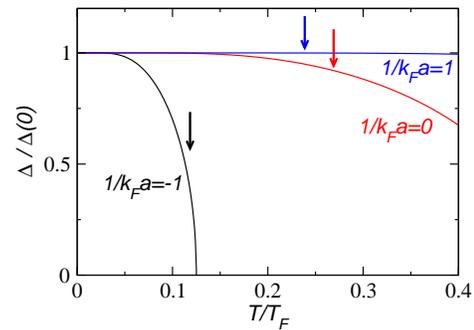}}
\caption{This figure shows the behavior of the excitation gap as a function
of temperature \cite{caveat} for the Fermi gases at three different scattering lengths.
This should be compared with Figure \ref{fig:7} for the cuprates. Arrows indicate locations of $T_c$.}
\label{fig:5ff}
\end{figure}

\begin{figure}
\includegraphics[width=2.4in,clip]
%{PEgdel2.eps}
{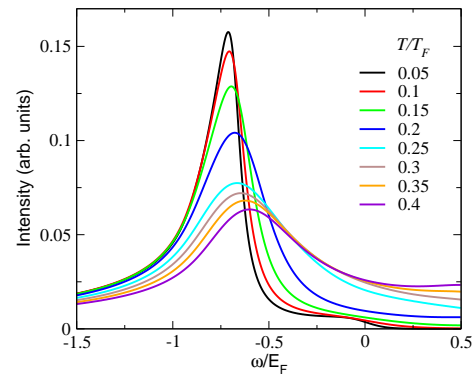}
%{PEg0.5s0.1.eps}
\caption{This is a photoemission-like plot for a homogeneous unitary Fermi gas
based on
Eqs.~(\ref{SigmaPG_Model_Eq})
and
(\ref{eq:photo1}).
Here $T_c = 0.27$ and $T^*\approx 0.5T_F$. The figure shows that the onset of superfluid
coherence leads to a sharpening of the peak structure. We take
$\gamma$ to be $0.5E_F$. This figure can be compared with
Figure \ref{fig:7}.}
\label{fig:6ff}
\end{figure}
   
\begin{figure}
\centerline{\includegraphics[clip,width=2.5in,height=3.0in]
{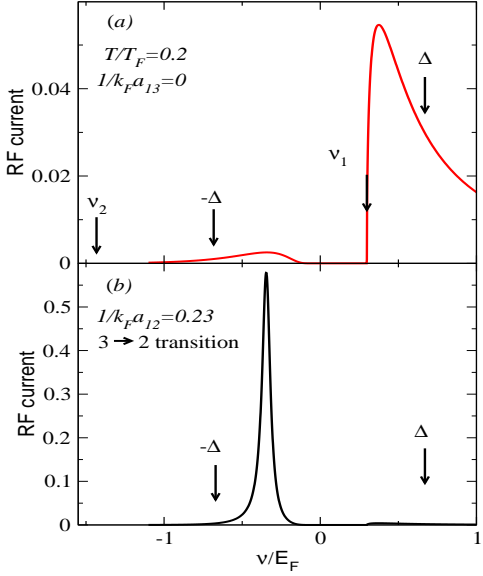}}
\caption{Pedagogical figure showing typical RF spectra of unitary
homogeneous gas at temperature somewhat below $T_c$. The various characteristic
energy scales are labelled. Upper panel corresponds to
absence of 
final state effects while lower panel includes final state effects in
rather extreme limit of a (weakly) bound
state in the negative continuum. This is optimal situation for using the
sum rules to extract $\Delta$.}
\label{fig:7ff}
\end{figure}

\begin{figure}
\centerline{\includegraphics[clip,width=2.5in,height=2.1in]
{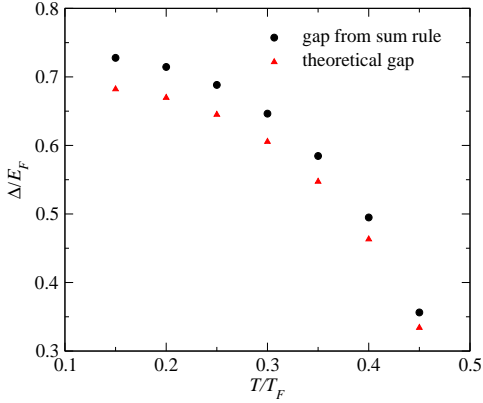}}
\caption{Based on previous figure, plotted here are pairing gaps
$\Delta(T)/E_F$ vs temperature as computed exactly and as computed from
a limited integration, using the sum rule.}
\label{fig:8ff}
\end{figure}

We present in 
Figure \ref{fig:5ff} a plot of the excitation gap in the cold gases
for three different values of the $s$-wave scattering length
in units of $1/k_Fa$ which are near unitarity ($a = \infty$)
and on both the BCS and BEC sides. This figure should be compared
with Figure \ref{fig:7} for the cuprates. Here the excitation
gap  is estimated
using Eq.~(\ref{eq:gap_equation}) for all temperatures \cite{caveat}.
%above $T_c$, although one should ultimately include finite
%$\mu_{pair}$ effects.
Also indicated on the curves is the value of the transition temperature.
This figure makes it clear that pseudogap effects, which are essentially
absent 
on the BCS side of resonance, are very apparent at unitarity, where the
Fermi gas has a positive chemical potential.
In both the unitary and BEC cases, $\Delta$ is roughly
temperature independent below $T_c$.

Figure \ref{fig:6ff} represents a photoemission-like
study, but for the parameters associated with a unitary (homogeneous)
Fermi gas.
Here
the vertical axis plots the ${\bf k}$ integral of
$I^{photo}({\bf k}, \omega)$ based on Eq.~(\ref{eq:photo1})
assuming a structureless matrix element $M_0$.
This figure should be compared with
the cuprate data in Figure \ref{fig:1}. 
The various curves correspond to different temperatures as indicated
with $T_c = 0.25T_F$ and $T^*\approx 0.5T_F$. 
The self energy is based on
Eq.~(\ref{SigmaPG_Model_Eq})
for the non-condensed pair component with $\gamma = 0.25E_F$ and
$i\Sigma_0 = \Gamma_0 = 0.1E_F$. 
What is most notable about this figure is the progressive sharpening of the
``photoemission" peaks associated with the growth of coherence
as $T$ decreases. This
same effect is seen in the cuprate data
(Figure \ref{fig:1}).  
One notes here, however, that there is some shift of this
peak position reflecting an increase in
$\Delta$ with decreasing $T$, which is not seen in the
cuprate data. This effect can be attributed
to the fact the there is a substantially larger
separation \cite{MicnasRMP,Chienlattice,Chienlattice2}
between $T^*$ and $T_c$ in the case of a lattice
(away from the BCS regime) than for a gas such as shown here.
This is apparent in Figure \ref{fig:40a}.
Thus, there is more temperature dependence found in
the excitation gap of the superfluid
phase (if one compares with the same value of
$T^*$).

\begin{figure}
\includegraphics[width=2.5in,clip]
{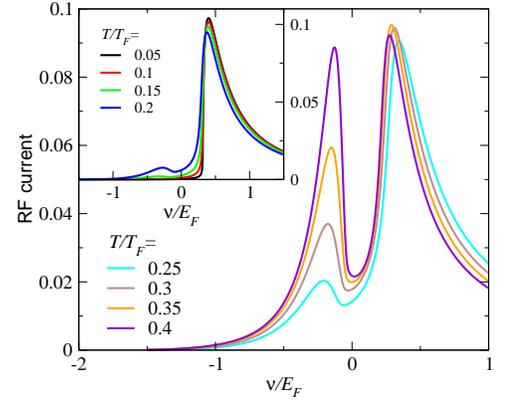}
%{rfg0.5s0.eps}
\caption{This is the RF counterpart of Figure \ref{fig:6ff} for
homogeneous unitary gas. The main
body of the figure plots the higher $T$ behavior and the inset shows the
results at lower $T$ when superfluid coherence is well established. Just
as in photoemission, there appears to be a signature of this coherence
in the RF spectra which is associated with a rather sharp threshold
behavior, as seen in the inset.}
\label{fig:9ff}
\end{figure}

\begin{figure}
\includegraphics[width=2.5in,clip]
{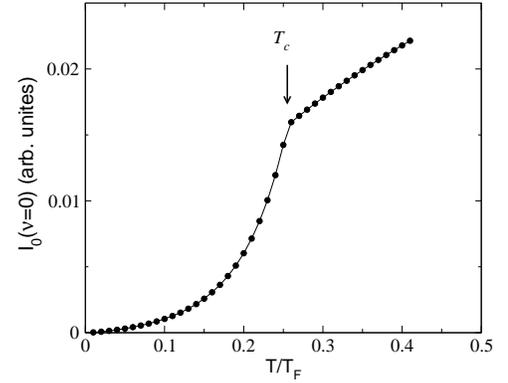}
\caption{Evidence that $T_c$ may show up as subtle feature in
RF spectroscopy. Based on the results of 
Figure \ref{fig:9ff}, 
plotted here is the RF current magnitude at zero detuning (for a homogeneous
balanced unitary gas without final state effects) as a function of temperature.}
\label{fig:10ff}
\end{figure}

\subsection{Overview of RF spectra: Homogeneous Case}

The top panel of Figure \ref{fig:7ff}
represents a plot of characteristic RF spectra for a unitary
gas without final state effects
and at a moderate temperature below $T_c$.
Here we use strict mean field
theory. Indicated in the figure are the various energy scales
showing the location of the pairing gap $\Delta$
as well as the
thresholds associated with the
negative and positive continua.  
Spanned by $-(\sqrt{\mu^2 + \Delta^2} +\mu) \le \nu \le 0$ 
and $\nu \ge \sqrt{\mu^2
  + \Delta^2} -\mu $ these are indicated as
$\nu_2$ and $\nu_1$. 
One can see that there is a substantial separation between
the pairing gap value and the 
threshold $\nu_1$. Thus, there is very little in
the figure to suggest a way of extracting the pairing gap.
This has presented a dilemma for the field.

One way to address this issue is to exploit the sum rule
in
Eq.~(\ref{eq:14}) which is appropriate provide one includes
final
state effects. 
In the lower panel we show the same spectra when final state
effects are
included. We have chosen a very special case for
illustrative purposes in which a (meta-stable) bound state
overlaps the negative continuum. This represents the most
ideal example for exploiting sum rule constraints
to extract the pairing gap. 
One can see here that because the bound state is in the 
negative continuum, the bulk of the spectral weight is
confined to a narrow frequency weight spanning from
$\nu_2$ to $0$.

In Figure \ref{fig:8ff}
we show the estimated values for the pairing gap $\Delta$ of
a unitary gas obtained
from the sum rule as integrated from $\nu = -2E_F$ to $\nu = +2E_F$
compared with the exact pairing gap. 
The accuracy is within 10 \%. 
To arrive at a case where the final state is on the
BEC side of resonance is reasonably straightforward and the
13 superfluid, which exhibits this,
is now well studied by the MIT group \cite{Ketterlepairsize}.
However, we point out that for this 13 superfluid and
for typical values of $k_F$ the
bound state is deep and well removed from the continuum. By contrast, the case
shown here results from a situation in which $k_F$ is increased
from the currently quoted experimental values by about a factor of 10.
While this may not be easy to achieve in the near future, it does point
to the advantage of exploiting final state effects to focus the
spectral weight in the more well confined, negative $\nu$ regime.

We plot in Figure \ref{fig:9ff}
the homogeneous spectra in the absence of final state effects
but now for the case in which we go beyond strict mean field
theory and, thereby, differentiate the condensed and non
condensed pairs on the basis of 
Eq.~(\ref{SigmaPG_Model_Eq}). We have chosen the same parameters
as in 
Figure \ref{fig:6ff}. In contrast to this earlier photoemission-based
calculation one 
applies
Eq.~(\ref{RFc0})
to describe the RF experiments.
Each of the curves corresponds to the same temperatures as their
counterparts in 
Figure \ref{fig:6ff}. It can be seen that the shape of the
RF spectra is very different from that of photoemission
(even in the absence of final state effects), because of the
constraint 
introduced by $\omega=\ek -\delta\nu$ which appears in
Eq.~(\ref{RFc0}). There are two peak structures at higher $T$,
even in this homogeneous situation, with the lower peak
corresponding to the negative continuum.
The temperature regime is separated into the superfluid phase in
the inset and the normal phase in the main body. The differences
between the two sets of curves supports the fact that
there are signatures of superfluid coherence in RF spectroscopy.
While, there are no abrupt changes at $T_c$, nevertheless,
like the quasi-particle peak sharpening seen in the cuprates (in
Figure \ref{fig:1}) one sees a distinctive sharpening of
the positive $\nu$ threshold as the temperature crosses
$T_c$.

To quantify this, Figure \ref{fig:10ff} shows a plot of the zero detuning current
for this balanced gas as a function of temperature. There is
a feature at $T_c$ which may make it possible, in principle, to
extract this transition temperature from high resolution RF experiments.
This feature reflects the fact that there are subtle but
distinctive features in the RF spectrum associated with superfluid
coherence, as seen in
Figure \ref{fig:10ff}. Indeed, one infers from this previous
figure that the negative continuum only begins to be perceptible
as the temperature approaches $T_c$.

\subsection{Momentum Resolved RF}

\begin{figure}
\includegraphics[width=2.5in,clip]
{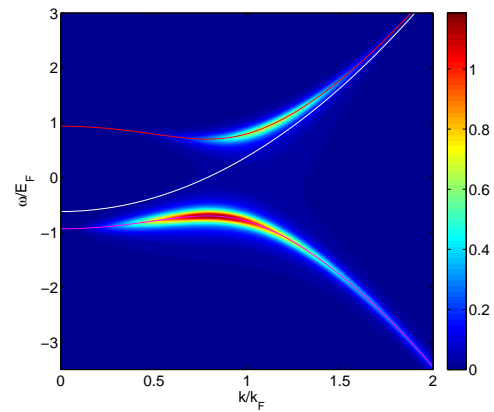}
%{InvkFa0K100lmSg0.02gm0.02R0.22T0.26Akk2_contour.eps}
\caption{
%(a) Energy levels in an RF transition.
Contour plots of momentum resolved RF spectra for the homogeneous
case at unitarity at $T/Tc \approx 1.9$. The top and bottom lines in the
contour plots correspond to the two quasiparticle dispersions
$\mu$ $\pm \sqrt{(k^2/2m-\mu)^2 +\Delta^2}$ and the middle line is
that of the free atom
dispersion. } \label{RFm}
\end{figure}

We turn now to momentum resolved RF spectra.
Later in
Figure \ref{Fermiarc},
we will
study some potentially analogous experiments which address
momentum resolved (or ``angle resolved") photoemission in the cuprates.
In Figure \ref{RFm}, we present an intensity map
as a function of single particle energy and
wave vector $k$ for the (occupied) states which contribute to the
RF current. Here we consider the homogeneous case.
The yellow to red regions of the figure correspond to where the
spectral intensity is highest.
The
temperature here is chosen to be relatively high, around $1.9 T_c$
in order to have some contribution to the RF current from
pre-existing thermal fermionic excitations. The intensity map
indicates upward and downward dispersing contributions. These
correspond, to a good approximation, to the two RF transitions:
to state 3 from state 2
with dispersion ($\Ek + \mu$) and ($-\Ek + \mu$), respectively,
which are shown in Figure \ref{RFtran}. 
These are 
measured relative to the bottom
of the band.
The width of this contour
plot comes exclusively from the incoherent terms $\gamma$ and
$\Sigma_0$. Here we have chosen to represent the latter by
an imaginary constant (as in the cuprate literature)
$\Sigma_0 = - i \Gamma_0$.
Based on
Eq.~(\ref{SigmaPG_Model_Eq}) and for
illustrative purposes we
take $\gamma = 0.25 T_F$ and $\Gamma_0 = 0.1 T_F$,
with a small resolution broadening (typical of the experiment)
as well.

One can see that the bulk of the current even at this
high temperature is associated with the pair states
which are broken in the process of the RF excitation. This figure
describes in a conceptual way, how this intensity map can be used
to compare with a broadened BCS-like form for the spectral
function. This form fits very well the two branches shown in the
figure corresponding to upward and downward dispersing curves.
In this way one can, in principle, establish the presence
of pairing and extract the pairing gap size.

We stress that these calculations are for the homogenous case
and it is important to extend them to include the effects of a
trap. This can be done within an LDA approximation scheme. Once the
trap is included the simple analogy between the electronic ARPES
experiments and momentum resolved photoemission spectroscopy is
invalidated. However, many of the central features survive. While
the two branches shown in the contour plot in Figure \ref{RFm}
are, in principle present, there is a third new branch which
appears as well. This corresponds to essentially free atoms at the
trap edge which will contribute significantly \cite{Torma2,heyan}
to the RF current.
It is this branch which is also upward dispersing which makes it
rather difficult to see the effects of the pre-existing thermally
broken pairs.

We summarize the results shown elsewhere \cite{ourRF4} for
the behavior of a unitary trapped gas over a range of
different temperatures.
At
high $T$,
the central notable
feature is a single upward dispersing
curve which fits the free particle dispersion.
This dispersion can be readily differentiated from that
associated with pre-existing thermally broken pairs
which varies as
$\Ek + \mu$
and of course, depends on the distribution of energy gaps
$\Delta(r)$.
It arises from
free atoms at the trap edge
(where the gap $\Delta(r)$ is small).
As the temperature is decreased towards $T_c$ a second
(downward dispersing) branch becomes evident. In the vicinity
of the transition, the intensity
map is
bifurcated with two co-existing peaks: one coming
from the free atoms at the trap edge
and the second from the condensate pairs which are broken
in the process of the RF excitation.
The separation of the two peaks can be difficult to discern until
sufficiently high values of $k$.
Finally, at the lowest temperatures the striking
feature is a single
downward dispersing branch.
This reflects the fact that essentially all atoms are now paired in the condensate.
Just as in the homogeneous case discussed above, a BCS-like fit
to this dispersion can be used to determine the pairing gap.
We stress that there are no abrupt changes in the RF behavior
at the superfluid transition, very much like what we saw
earlier in our summary of the cuprate literature.

\begin{figure}
\includegraphics[width=2.5in,clip]
{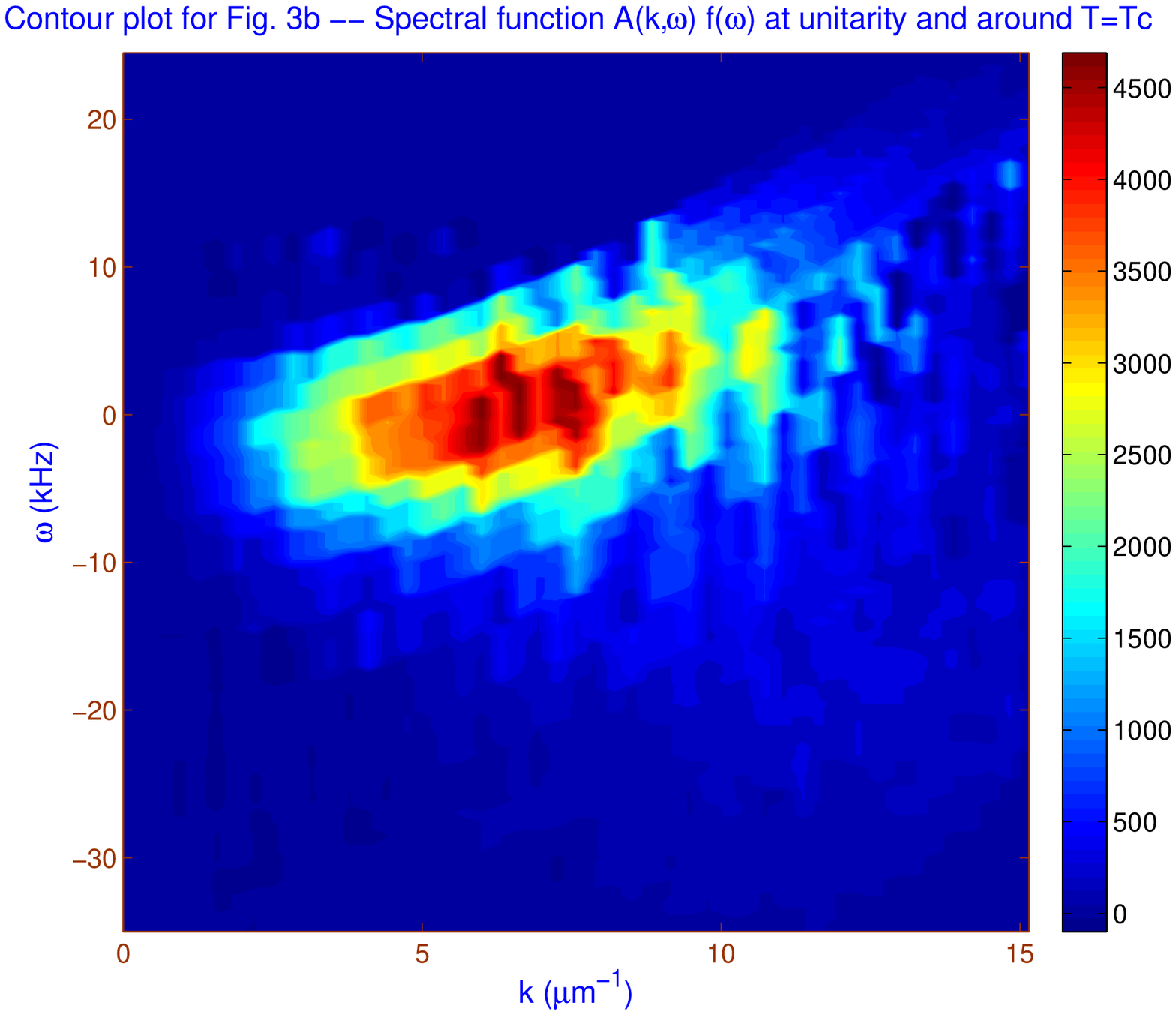}
\includegraphics[width=2.3in,clip]
{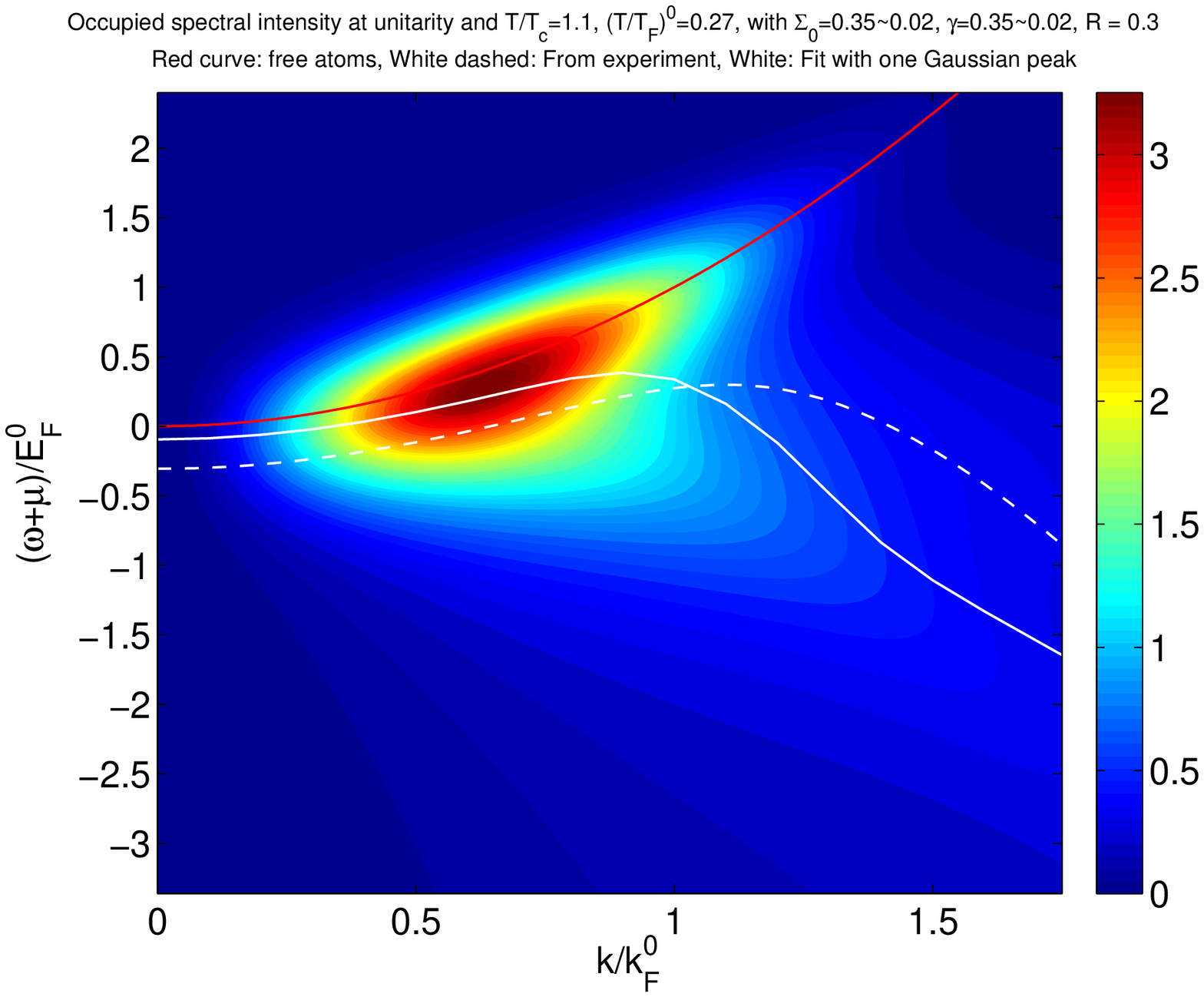}
%{UnitarySg0.35-0.02gm0.35-0.02R0.3T1.1contour_2.eps}
\caption{ Contour plots of momentum resolved RF spectra in a
trapped configuration. Top panel (a) is experimental data
\cite{Jin6}.
Theoretical results (b)
correspond to occupied
spectral intensity map, in a unitary trapped Fermi gas at $T/T_c =
1.1$. Here $\Sigma_0 = 0.35 E_F^0$ and $\gamma = 0.38 E_F^0$ at the
trap center.
%, and WHAT IS THIS $\sigma = 0.3 E_F^0$.
The red curves represent the free atom
dispersion, while the white solid and dashed curves are the
quasiparticle dispersion obtained theoretically and experimentally
\cite{Jin6}, respectively, via fitting the energy distribution
curves (EDCs) with a single
Gaussian.} \label{RFmEXP}
\end{figure}

\begin{figure}
\centerline{\includegraphics[clip,width=2.5in,height=2.8in]{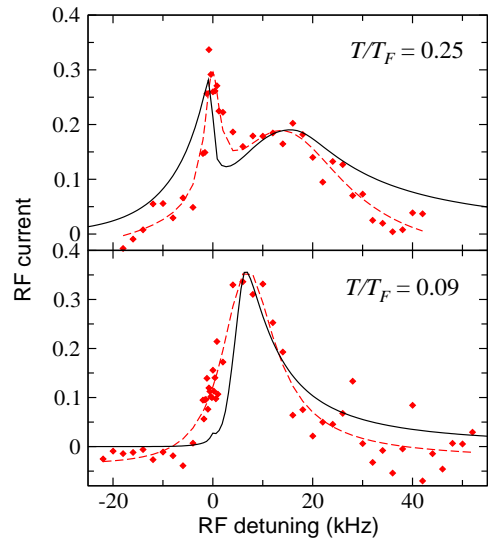}}
\caption{Comparison of calculated RF spectra of a trapped gas (solid curve, $T_c
\approx 0.29$) with experiment \cite{Grimm4} (symbols) in a harmonic trap
calculated at 822 G for two (estimated) temperatures. From Ref.~\cite{heyan}.
The dashed lines
are a guide to the eye. There is reasonable agreement, but
because final state effects are not included, the high frequency
tails are overestimated in the theory.} \label{RFtrap}
\end{figure}

\begin{figure}
\includegraphics[width=2.1in,clip]
{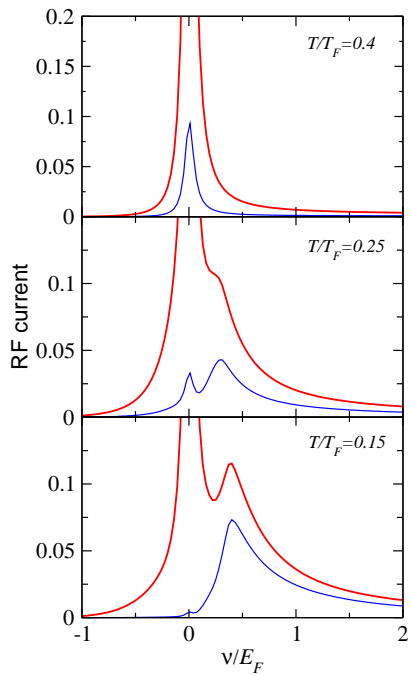} \caption{RF spectra for a trapped unitary imbalanced gas
with $\delta = 0.5$. Here we take $\gamma = 0.05$. Four different
temperatures are indicated.
Majority spectra are in
red and minority in blue.} \label{RFtrapImb}
\end{figure}

\section{Analysis of Theoretical and Experimental RF Spectra}

\subsection{Momentum Resolved Spectroscopy}

We now compare theory and experiment in a trap based on the
momentum resolved spectra just discussed for the homogeneous case.
In Figure \ref{RFmEXP} we have taken a  larger intrinsic
broadening and included an empirical resolution broadening as
well, again somewhat larger than the value $R \approx 0.2$
indicated for the experiments. These parameters are seen to
optimize semi-quantitative agreement with the data plotted in the
top panel from Reference \cite{Jin6}.

The bottom panel presents the theoretical intensity maps.  The
dotted white curve represents a fit of the experimentally
deduced peak dispersion while
the solid white curve is the theoretical counterpart.  Here, as in
experiment we have fit the energy distribution curve to a
\textit{single} Gaussian peak. The comparison between the two
white curves shows semi-quantitative consistency. Moreover, both
the solid and dotted white curves can be well fit to the BCS
dispersion involving $\Ek$, as was originally proposed in Ref.
\cite{Janko}. While Figure \ref{RFmEXP} seems to capture
the essential results shown in the experiment, with higher
resolution it should be possible to obtain more direct information
about the mean experimentally-deduced gap size.
Importantly, this reasonable agreement and the fact that
the experiments were done near $T_c$
suggests that there is a sizable pseudogap in the Fermi
gases at and above $T_c$ at
unitarity.

\subsection{RF Spectra in a Trap}

In Figure \ref{RFtrap} we compare RF spectra in a trap near unitarity
(solid curve) with experiments from Reference \cite{Grimm4}
(symbols) at 822~G on $^6$Li and
for two different temperatures. The dashed curve is a fit to the
data, serving as a guide to the eye. The higher $T$ curve indicates a
two peaked structure in both data and theory. The lower peak was
shown \cite{heyan} to be associated with nearly free atoms at the
edge of the trap. The upper peak reflects the existence of
pairing. Although it is not possible to directly infer the size of
the (trap averaged) pairing gap $\Delta$, it is now reasonably
clear \cite{Varenna} that a pairing gap (pseudogap) is present in
the normal state even in these early experiments from the Innsbruck
group. The lower curve can be interpreted to suggest
that the atoms at the trap
edge have temperature small compared to $\Delta(r,T)$. The
agreement between theory and experiment is not unreasonable for
this leading order calculation (based on $I_0(\nu)$). One can,
however see that the theory in both cases shows a much slower
drop-off with increasing high frequency, than seen experimentally. We will see
shortly that this difference is associated with final state
effects.

In Figure \ref{RFtrapImb} we present similar RF spectra for
\cite{heyan2} a trapped imbalanced gas near unitarity. The
polarization is $\delta = 0.5$, and the spectra are plotted for
three different temperatures. Here $T_c/T_F =0.25$. 
It is useful to refer back to the lower right panel in
Figure \ref{fig:40b} to see precisely what region of the 
polarized gas phase diagram
is relevant. 
The red curves
correspond to the majority and the blue to the minority. For the
majority, one can see that the free atom peak at $\nu =0$ is
present at all temperatures, unlike the previous case in a balanced
gas. At the highest temperature $T/T_F = 0.4$ (which is close to
$T^*\approx 0.35$) the system is normal and pairing is absent. Very close to
$T_c$ in the middle panel one sees a clear pairing peak signature
associated with the pseudogap. For this analysis we chose the
broadening in Eq.~(\ref{SigmaPG_Model_Eq}), to be very small
with $\gamma=0.05E_F$.

\subsection{Final State Effects}
\label{FinalStateCMP}

\begin{figure}
\includegraphics[width=2.5in,clip]
{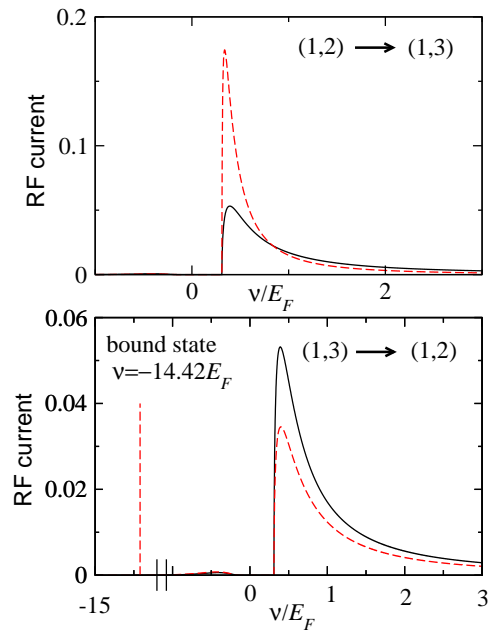} 
\caption{Comparison of homogeneous RF spectra of
a unitary gas with (red,dashed) and without (black)
final state effects at $T/T_F = 0.15$. The top figure is for 12 superfluid
and the bottom is for 13 case, showing a bound state. Calculations were done
with $\gamma =0$.}
\label{ChinMIT}
\end{figure}

\begin{figure}
\includegraphics[width=2.0in,clip]
{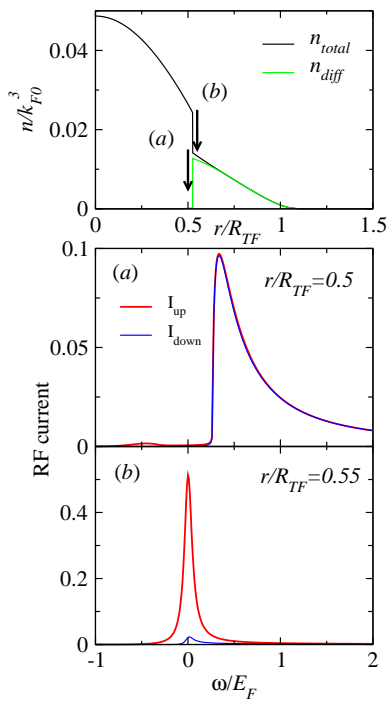} \caption{Homogeneous RF
spectra of unitary, population imbalanced ($\delta = 0.5$)
gas at very low $T=0.05$, where phase
separation is stable. Upper panel is density profile indicating radii (a) and (b)
used in lower two panels to compute the tomographic or homogeneous
spectra. Red is majority and blue is minority. No final state effects are
included and we take $\gamma =0.05$.}
\label{RFtomoLowT}
\end{figure}

\begin{figure*}
\includegraphics[width=5.0in,clip]
{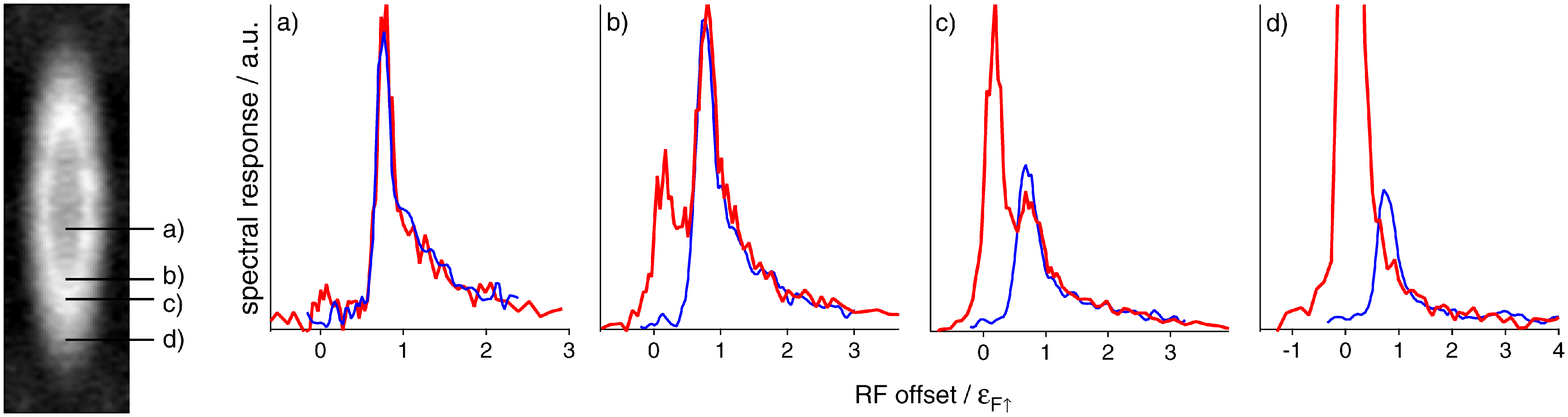}
\includegraphics[width=5.5in,clip]
{rfT0.15a0p0.5cmp.eps} \caption{Comparison between theory and
experiment \cite{MITtomoimb} in top panel, for tomographic scans of an imbalanced
unitary gas. Theory assumes $\delta = 0.5$, for definiteness.  Here
$T/T_F = 0.15$. The calculations are for a 13 superfluid (as in
experiment) with final state effects included
($1/k_Fa_{12}=2.5$), and $\gamma =0$.
Red indicates majority
and blue, minority. Panels a, b and c in the theory should be compared
with b, c and d in experiment.} 
\label{RFtomoMIT}
\end{figure*}

As we saw in Figure \ref{RFtrap}, final state effects are expected
to cut off the long tails in the RF spectra found theoretically in
the lowest order theory. It has been argued
\cite{Ketterlepairsize} by the MIT group that one should limit the
importance of these final state contributions by studying a 13
superfluid instead near unitarity with an RF transition involving
(for example) $ 3 \rightarrow 2$. For this case the magnitude of the final state
scattering length is small, although it is positive and this will
lead to a bound state contribution in the spectra.
One could alternatively argue that it is better to work with
the 12 superfluid where there are (generally) no bound states and
where one can, more readily, impose the sum rules to arrive at estimates of the
pairing gap.  At this point both options should be explored.

In Figure \ref{ChinMIT} we compare homogeneous spectra at
$T=0.15T_F$ with and without final state effects for these two different
superfluids at unitarity. The figure on the top corresponds to
the configuration of the Innsbruck experiments \cite{Grimm4} and
on the bottom to recent MIT experiments \cite{Ketterlepairsize}. It
can be seen that final state effects in both cases do not change
\cite{YuBaym} the threshold $\approx 0.4\Delta$ for the positive
continuum (discussed in Section \ref{FinalStateTheory}). However,
they do lead to a somewhat sharper peak and to a more rapid fall
off at high detuning, as is consistent with the sum rules.

It is not easy to do the calculations which include final state
effects in a trap so we can only qualitatively infer from the top panel
in Figure
\ref{ChinMIT} that the corrections associated with their
inclusion are what is needed to improve the agreement
between theory and experiments on the 12 superfluid in Figure
\ref{RFtrap}. 
If final states introduced a bound-bound transition on the BCS side of
resonance, as conjectured \cite{Ketterlepairsize}, it would probably
not be sufficiently robust with respect to temperature \cite{ourRF4},
and would likely merge with the continuum with increasing $T$. Moreover, this 
positive continuum contribution, \textit{which reflects pairing}, 
is always present; theoretically, one never finds
simply an isolated bound-bound transition. With a small amount
of lifetime broadening ($\gamma$) it is likely that a final
state induced bound-bound transition on the BCS side
of resonance would not make a noticeable
difference.
In this way, although some concern has been
raised \cite{Ketterlepairsize} about whether or not the
Innsbruck experiments were properly interpreted as evidence
for a pairing gap (rather than a possible bound-bound \cite{Basu} transition),
we concur with their original interpretation.

By contrast with the 12 superfluid, when final state effects
for the 13 superfluid are included, there is little change in the
shape of the spectra (shown
on the bottom).
There is, however,
an important change of
vertical scale associated with the bound state.

\subsection{Tomographic Scans in Imbalanced Gases}

In Figure \ref{RFtomoLowT} we turn to effectively
homogeneous spectra associated with tomographic plots in the
same imbalanced gas with $\delta = 0.5$ studied
earlier. Here we consider extremely low
temperatures so that \cite{ChienPRL} the system is in the phase separated state
as can be seen from the lower right plot in Figure \ref{fig:40b}.
One can see this phase separation in the top panel which presents the
density profiles. Also indicated are the two points (a) and (b) which
establish the radii used in the lower panels at which the RF
spectra are plotted. Here $R_{TF}=\sqrt{2E_F/m\omega^2}$ is the Thomas-Fermi radius
and $m$ and $\omega$ denote fermion mass and trap frequency. The red and blue curves
are for the majority and minority. The middle panel shows
that just inside the superfluid core there is very little
difference in the majority and minority spectra as expected for
a
locally unpolarized superfluid. The lower panel shows the behavior
just on the other side of the phase separation boundary where there is very
little minority and hence very little pairing. As a consequence
the RF peaks are close to zero detuning.
We stress here that we are using the strict BCS-Leggett
theory without including Hartree effects. With the
latter included there may be a relative shift of the
energy scales associated with the majority and minority
atoms.

Figure \ref{RFtomoMIT} presents a comparison between theory (with
final state effects) and experiment \cite{MITtomoimb}
at moderate temperatures ($T_c = 0.25T_F$) within the Sarma
phase. The theory and experiments are for the 13 superfluid at
unitarity.
The upper panel corresponds to recent
data from MIT \cite{MITtomoimb}, indicating via a contour plot, the various radii
probed in the tomographic scans.  In the lower panel, the
counterpart theoretical profile indicates the three different
radii via (a), (b) and (c). Also shown in the theory is where $-
\Delta$ would be found within the negative continuum. One sees a
reasonable correspondence between theory and experiment, if we compare
b,c and d on the top with (a), (b) and (c), respectively on the bottom. We 
stress that Hartree effects have not been included in the
theory so that the zeroes of the
horizontal energy scales are not equivalent.
These Hartree effects have been extensively analyzed in Ref.~\cite{MITtomoimb},
and these authors, as well, have provided estimates of the pairing gap.

\section{Photoemission Experiments in the Cuprates}
\label{PhotoEXP}

 We turn now to recent issues in photoemission
experiments \cite{Kanigel} in the underdoped cuprates. These call attention to
the question of how the $\bf{k}$ dependence of
the spectral function varies as one crosses
$T_c$. The earlier discussion in
Figure \ref{fig:1} focused on the ${\bf k}$ integrated photoemission
spectra. As in this previous figure here we address how 
superconducting coherence is manifested when
there is a normal state pseudogap. We stress that measurements like
photoemission and RF spectroscopy are not phase sensitive probes of
the system and cannot directly prove the existence of
superfluidity. 

Here we
focus on the regime
near the gap nodes, where the gap is smallest.
The experiments of interest here very likely contain important clues
as to the nature of the superconducting state which appears
in the presence of a normal state pseudogap. While many aspects
of the cuprates below $T_c$ appear to be typical of ($d$-wave)
BCS superconductors one expects some differences to appear simply
because an excitation gap is present at the onset of superconductivity.

There are four key points which have been identified
in Ref.~\cite{Kanigel}. We believe these are consistent with
a BCS-BEC crossover interpretation of these cuprate
photoemission data.
As reported \cite{Kanigel} (i) the
excitation gap, $\Delta({ \bf k})$, as measured in photoemission
experiments remains roughly constant in temperature from very low $T$,
to temperatures well above $T_c$.  (ii) In the superfluid phase
$\Delta({ \bf k})$ displays the expected point nodes (associated with
$d$-wave symmetry); however, these rapidly broaden into Fermi arcs once
the temperature reaches the vicinity of, and surpasses $T_c$.
Importantly, ``this remarkable change occurs within the width of the
resistive transition at $T_c$''. (iii) It has also been reported
\cite{KanigelNature} that the energy scale associated with the
excitation gap appears to be $T^*$, which is conventionally taken as the
pseudogap \cite{LeeReview,ourreview,Varenna} onset temperature, and that the Fermi
arc length scales with $T/T^*$ above $T_c$.  From (i) it is inferred
that (iv) ``the energy gap is \textit{not} directly related to the
superconducting order parameter''.

To address these and other photoemission experiments, the normal
state self energy is taken \cite{Normanarcs,Chubukov2,Chen4}
to be of the form shown in
Eq.~(\ref{SigmaPG_Model_Eq})
with 
Eq.~(\ref{selfE3})
and with a purely imaginary background self
energy: $\Sigma_0 ({\bf k}, \omega) = -i \Gamma_0$.
The rapid, but smooth destruction of the $d$-wave point
nodes can be physically associated with
the fact that the superconducting order parameter $\Delta_{sc}$
disappears smoothly but precisely at $T_c^+$.  Above $T_c$ the
effects of $\gamma$ and $\Sigma_0$ lead to a smearing 
and the point nodes are replaced by Fermi arcs \cite{Normanarcs}.
Below $T_c$ with the onset of phase coherence through $\Delta_{sc}$,
the arcs are rapidly
replaced by point nodes. One says that there has been 
"a collapse of the Fermi arcs", and that the nodes are "protected" below
$T_c$.

The collapse is a
continuous process.
We argue that it
is not to be
associated with an abrupt disappearance of the inverse 
lifetime $\gamma$, but rather it reflects the gradual emergence of
the condensate to which the finite momentum pairs are continuously
converted as $T$ decreases.
This is related to the fact that, from
Eq.~(\ref{selfE3}), 
we see there are two terms in the self energy below $T_c$.
At the lowest temperatures $\Delta_{pg}$ vanishes, whereas above
$T_c$, it follows that $\Delta_{sc}$ is zero.

\begin{figure}
\centerline{ \includegraphics[width=2.9in,clip] {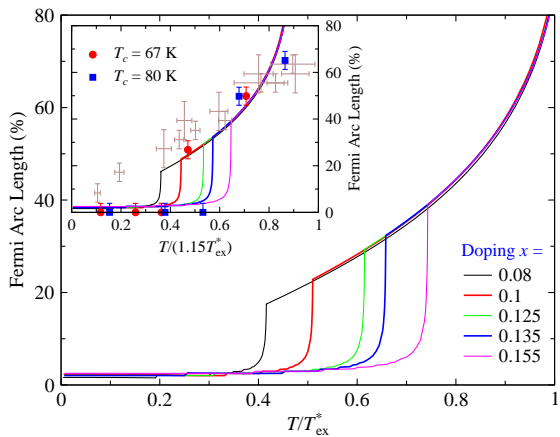} }
\caption{(color online) Fermi arc length as a function of
$T/T^*_{\mbox{ex}}$ for doping concentrations from optimal to
underdoping for a cuprate superconductor. Fermi arc length is
typically finite above $T_c$ and drops to zero upon the onset of
phase coherence. The normal state portions of the curves is close
to universal, in agreement with Ref.~\cite{KanigelNature}.
The comparison in the inset between the theory with a slightly
(15\%) enlarged $T^*_{\mbox{ex}}$ and experimental data (symbols)
\cite{Kanigel} shows a good semi-quantitative agreement.}
\label{Fermiarc}
\end{figure}

In Figure \ref{Fermiarc} we address
these new experiments by showing the
collapse of the Fermi arcs from above to below $T_c$ within
our BCS-BEC crossover formalism; we
plot the percentage of Fermi arc length as a function of $T/T^*$
and for different doping concentrations from the optimal ($T^* \approx T_c$)
to the
underdoped regime ($T^* >> T_c$).
The observed collapse is intimately
connected with our earlier observation that
the spectral function in Eq.~(\ref{spec}) has a zero at
$\omega=-\xik$ below $T_c$, whereas the spectral function
has no zero above $T_c$.

In summary, this reasonable agreement between theory and experiment
shown in the plot supports
our physical picture that pseudogap
effects persist below $T_c$ in the form of noncondensed pair
excitations of the condensate.  
We argue that the
pre-formed pairs above $T_c$ do not abruptly disappear just below
$T_c$.
This recognition that the
superfluid phase is, itself, very complex represents a new
direction for the cuprate field which these important ARPES
experiments have now called to our attention.
We also note that this purportedly more complex superfluid, containing non-condensed
pair excitations can potentially be elucidated through studies of
BCS-BEC crossover in the cold Fermi gases.

\section{Conclusions}

There has been enormous progress in the field of radio frequency
(RF) spectroscopy of the Fermi gases. This technique
holds the promise of being as valuable to these atomic
superfluids as photoemission has been
to the cuprates. We have tried in this Review to
argue that it also holds the promise of helping to address
some fundamental issues in the cuprates which are very general, such as
how to describe that anomalous superfluid
phase which forms in the presence of a normal state excitation gap.

On a less general level, these RF experiments also hold the promise of helping to address
(that is, support or rule out) one particular approach to the
theory of high temperature superconductivity: namely that based
on BCS-BEC crossover.
There are many alternative physical pictures of the cuprates and, indeed,
the ultracold gases on optical lattices have presented themselves
as possible simulators of at least some of these alternatives. Most notable 
among these are proposed cold gas studies
of the repulsive Hubbard model which is thought to be
relevant to the ``Mott physics" aspects of the high $T_c$ superconductors.

A goal of this Review was to present a broad background to the theory of
RF spectroscopy and its relation to photoemission spectroscopy.
We summarize key experiments using both techniques and show how
they can be addressed within a BCS-BEC crossover approach.
Included in our analysis are trapped as well as homogeneous gases,
and population imbalanced gases. The RF field has seen a proliferation
of methodologies including tomographic and momentum resolved
scans, all of which are discussed here.

The immediate excitement surrounding these RF experiments is in large part because
they hold the potential for measuring the pairing gap $\Delta$.
As time passes, however, it has become progressively more clear that
extracting detailed quantitative information is increasingly difficult.
There is no clear signature in the RF spectra of $\Delta$ except to
establish whether it is present or not. This observation applies
to both balanced and imbalanced gases.
Here we summarize two methodologies
which might make it possible to extract more precise
numerical values for $\Delta$:
either through sum rules which involve integrals of the measured
RF spectra, or through recent momentum resolved experiments.
At this moment, however, neither of these has been put into 
experimental practice
at a quantitative level.

Our analysis
(and all results presented in this
paper) are subject to the caveat that they are obtained within
the BCS-Leggett crossover theory (extended to finite $T$). This theory,
like all others is not an
exact theory.
However, this approach is
one which is particularly well suited to the cuprates because there are no spurious
first order transitions.
Recently, there have been RF calculations based on
an alternative scheme which is closer to the Nozieres Schmitt-Rink approach to BCS-BEC crossover.
This is very nice work, which we only have only briefly mentioned,
which addresses radio frequency experiments
in the normal state and at $T \approx 0$. The latter, in particular has
been important in elucidating how to incorporate
final state effects.
We also only make passing reference to the body of work
on the highly imbalanced gases.

It should be clear that this field is moving rapidly and much has been
accomplished at the theoretical and experimental ends. However,
in writing this Review, we felt it was timely to suggest possibly new directions
for cold gas research which might elucidate the cuprate superconductors.
As summarized earlier, the key issues which have emerged in photoemission
studies involve characterization of the fermionic self energy, of
the pseudogap and of the signatures of superconducting coherence (in passing
from above to below $T_c$).
These issues have a counterpart in the cold Fermi gases
as we have suggested here and it will be of enormous
benefit in future to exploit this new class of ``materials"
to address these fundamental questions in
condensed matter.

This work is supported by NSF PHY-0555325 and NSF-MRSEC Grant
0820054. We thank Cheng Chin for helpful conversations.

\appendix

\section{Analytical Results for RF Spectra in Homogeneous, Polarized Gas}
\label{App:analy} 

\begin{figure}
\begin{center}
\includegraphics[width=3.3in,clip]
{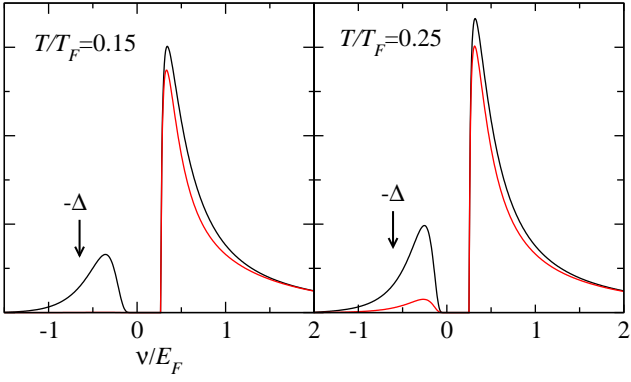}
\caption{RF current of the unitary polarized ($\delta=0.1$) gas at $T=0.15T_F$ 
and $T=0.25T_F$. For these temperatures, 
the Sarma state is stable. The black curve is for the majority and 
red is for the minority. The arrows indicate the position of $-\Delta$. $T_c\approx0.24T_F$ in this case.
\label{fig:app1}}
\end{center}
\end{figure}

Following the same arguments as in Section \ref{sec:analy},
for the polarized case, we have for the majority current    
\begin{eqnarray}
 I_0^{(1)}(\nu)&=&\frac{1}{8\pi^2}\frac{\Delta^2}{\nu^2}k_0[1-f(E_0+h)]\,,\quad (\nu>\nu_1)\\
I_0^{(1)}(\nu)&=&\frac{1}{8\pi^2}\frac{\Delta^2}{\nu^2}k_0f(E_0-h)\,,\quad
(\nu_2<\nu<0) 
\end{eqnarray} 
and minority RF current  
\begin{eqnarray}
 I_0^{(2)}(\nu)&=&\frac{1}{8\pi^2}\frac{\Delta^2}{\nu^2}k_0[1-f(E_0-h)]\,,\quad (\nu>\nu_1)\\
I_0^{(2)}(\nu)&=&\frac{1}{8\pi^2}\frac{\Delta^2}{\nu^2}k_0f(E_0+h)\,,\quad
(\nu_2<\nu<0) 
\end{eqnarray}
Here $\mu=(\mu_{\uparrow}+\mu_{\downarrow})/2$ is the 
average chemical potential,
and $h=\mu_{\uparrow}-\mu_{\downarrow}$ is chemical potential difference. Since $E_0+h$ is always positive,
the minority RF current is not associated with a
negative continuum at low temperatures. 
On the other hand, since generally $E_0-h<0$, the majority RF current
has a negative continuum. For strictly zero temperature, the 
negative continuum is located in the range  
$-h-\sqrt{h^2-\Delta^2}<\nu<-h+\sqrt{h^2-\Delta^2}$.
In Figure \ref{fig:app1}, we show the RF current for the stable Sarma state for small population imbalance $\delta=0.1$.
We note that the negative peak location is not at $-\Delta$.

\begin{figure}
\begin{center}
\includegraphics[width=3.3in,clip]
{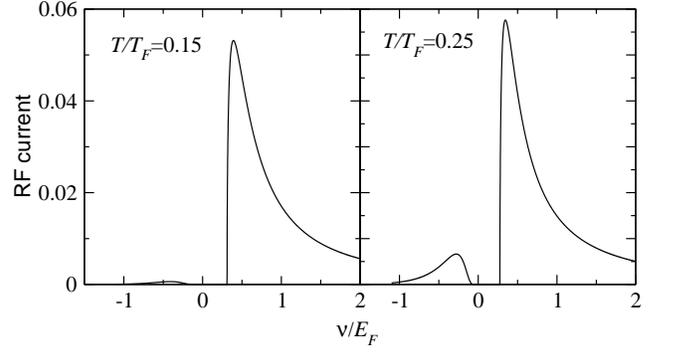}
\caption{RF current of the unitary unpolarized  gas at $T=0.15T_F$ and $T=0.25T_F$. 
 $T_c=0.255T_F$ in this case. This figure is included as a contrast with
 the polarized case, to make it evident that the negative continuum is
 more difficult to see in the absence of polarization.
\label{fig:app2}}
\end{center}
\end{figure}

In Figure \ref{fig:app2} we compare with the unpolarized case, showing
that there is essentially no negative continuum
(until one approaches temperatures closer to $T_c$).
We can ask why the onset of the negative continuum is unrelated to the energy
scale $\Delta$. 
The answer is that 
while, the
negative continuum in the Sarma phase is associated with the presence
of excess fermions (as seen by comparing the two figures in this appendix),
these excess fermions in the Sarma phase are essentially gapless.
Thus, they are not directly associated with an energy scale which involves
$\Delta$ in any explicit way.

\section{Details on the Final State Effect Diagrams}
\label{App:A}

The
response function which yields the RF current is \be D(\nu) = D_0(\nu) + D_2^2(\nu)\,
t^{}_{13}(\nu) \ee
where the first term is the usual leading order contribution and the second
corresponds to
the Aslamazov-Larkin (AL) diagram
given by

\ba & &D_{AL}(Q)=- \sum_{K,K',P}
G^{(2)}(K)G_0^{(1)}(-K)G_0^{(3)}(K+Q)\nonumber\\
&\times& t^{}_{12}(P) G^{(2)}(K') G_0^{(1)}(-K') G_0^{(3)}(K'+Q) t^{}_{13}(Q)  \nonumber\\
&=& \left[ \Delta  \sum_K G(K) G_0(-K) G_0^{(3)}(K+Q)  \right]^2
t^{}_{13}(Q)\nonumber\ea where, as discussed in the text we
have
$t_{12}^{} \approx -\Delta^2/T$. This term can be rewritten in terms of the functions
$F(Q)$, and $D_2{Q}$ defined in the main text.
We introduce the RF detuning frequency $\nu$
\be i\Omega_n
\rightarrow \Omega + i0^+ = \nu +\mu_2 -\mu_3 +i0^+ \quad . \ee

The
retarded form of the various complex functions which enter into the
RF current can be summarized as \ba
D_0^R(\nu) &=& \sumk \Big[ \frac{f(\Ek)-f(\xikk)}{\nu +\Ek -\xik+i0^+}\uksq\nonumber\\
&+&\frac{1-f(\Ek)-f(\xikk)}{\nu -\Ek -\xik+i0^+}\vksq \Big]\\
D_2^R(\nu) &=& \sumk \frac{\Delta}{2\Ek} \Big[\frac{1-f(\Ek)-f(\xikk)}{\nu-\Ek -\xik+i0^+}\nonumber\\
&-& \frac{f(\Ek)-f(\xikk)}{\nu  +\Ek -\xik+i0^+}\Big] \\
\chi^{R}_{13}(\nu) &=& -\sumk \Big[\frac{1-f(\Ek)-f(\xikk)}{\nu-\Ek -\xik+i0^+}\uksq\nonumber\\
&+&  \frac{f(\Ek)-f(\xikk)}{\nu +\Ek
-\xik+i0^+}\vksq+\frac{1}{2\epsilon_\mathbf{k}} \Big] \ea

To compute the RF current, we calculate the real and imaginary
parts of the above quantities. For later convenience, we introduce the
following expressions.  \ba B(\nu)  &=& \sumk \frac{1}{2\Ek} \Big[
\frac{1-f(\Ek)-f(\xikk)}{\nu-\Ek -\xik}
-\frac{f(\Ek)-f(\xikk)}{\nu  +\Ek -\xik} \Big]\nonumber\\
\label{Bf}\\
C(\nu) &=& -\pi \sumk \frac{1}{2\Ek} \Big\{ [1-f(\Ek)-f(\xikk)]\delta(\nu-\Ek -\xik)\nonumber\\
&-& [f(\Ek)-f(\xikk)]\delta(\nu +\Ek -\xik) \Big\}\label{Cf} \ea
We also define $B(\nu = 0) \equiv B(0)$ and
$C(\nu =0) \equiv C(0)$. For the most part it will be convenient
to drop the argument and simply write $B,C$.

It can be shown that \ba
\mbox{Im} D_0^R &=& \frac{\Delta^2}{\nu}  C\\
\mbox{Im}D_2^R &=& \Delta C \\
\mbox{Re}D_2^R &=& \Delta B \ea We notice that
\ba& &\chi^{}_{13}(\nu) - \chi^{}_{13}(0) \nonumber\\
&=& -\nu \sumk \frac{1}{2\Ek} \bigg[ \frac{1-f(\Ek)-f(\xikk)}{\nu
  -\Ek -\xik} - \frac{f(\Ek)-f(\xikk)}{\nu +\Ek -\xik}\bigg]
\nonumber\\
&=& -B\nu \ea We make use of the $1,2$ superfluid gap equation and
identify $\chi_{12}(0)=\chi_{13}(0)$, so that \ba
\mbox{Re}t_{13}^{-1,R} &=& \frac{m}{4\pi a_{13}} - \frac{m}{4\pi
a_{12}} +
\chi^{}_{13}(\nu) - \chi^{}_{13}(0)\nonumber\\
&=&\frac{m}{4\pi a_{13}}-\frac{m}{4\pi a_{12}}-B\nu\\
\mbox{Im}t_{13}^{-1,R}&=&\mbox{Im}[\chi^{R}_{13}(\nu) -
\chi^{}_{13}(0)]\nonumber\\ &=&-C\nu-B\,0^+ \ea

In the general case with $a_{12}\neq a_{13}$, it follows that \ba
\mbox{Im} D^R(\nu)&=&\mbox{Im}D_0^R\nonumber\\
&+&\frac{-\mbox{Re}(D_2^R)^2\mbox{Im}t_{13}^{-1,R} +\mbox{Im}
(D_2^R)^2\, \mbox{Re}t^{-1,R}_{13}} {(\mbox{Re}t^{-1,R}_{13})^2 +
(\mbox{Im}t^{-1,R}_{13})^2}\nonumber \ea After a straightforward
calculation we have \ba
\mbox{Numerator}=\frac{\Delta^2}{\nu}C\left(\frac{m}{4\pi
a_{13}}-\frac{m}{4\pi a_{12}}\right)^2 \ea

In this way we rewrite the response function in the form presented in the text
as
\ba \mbox{Im} D^R(\nu)=\left(\frac{m}{4\pi a_{13}} - \frac{m}{4\pi
a_{12}}\right)^2\frac{\mbox{Im}
D_0^R(\nu)}{|t_{13}^{-1,R}(\nu)|^2} \ea

\section{Special Case of Equal Interactions}

In the special case $a_{12}=a_{13}$, we expect
$I(\nu)
 \propto \delta(\nu)$.
We showed above that $\mbox{Im}t_{13}^{-1,R}=-C\nu-B\,0^+ $. The second
term plays essentially no role when $a_{12}\neq a_{13}$ case, but it is
important in this special case. Combined with the real part, we have
$$t_{13}^{R}=-\frac{1}{(B+iC)\nu+iB\,0^+}\,. $$
Using the fact that $C=0$ (when
$a_{12}=a_{13}$)
we find \ba
\mbox{Re}t_{13}=-\frac{1}{\nu} \frac{B}{B^2+C^2},\quad
\mbox{Im}t_{13}=\frac{1}{\nu}\frac{C}{B^2+C^2}+\frac{\pi}{B}\delta(\nu)\nonumber
\ea

Thus we have \ba \mbox{Im} D^R(\nu) &=& \mbox{Im}D_0^R(\nu)
+\mbox{Re} D_2^2 \,\mbox{Im} t^R_{13} + \mbox{Im}
D_2^{2,R} \, \mbox{Re} t^R_{13} \nonumber\\
&=& \pi \Delta^2 B\,\delta(\nu)\,. \ea
In the limit
$a_{12}=a_{13}$ \ba \Delta^2B(0)
&=& - \sumk \big\{ [ 1-f(\Ek)-f(\xikk)]\vksq\nonumber\\
&+& [f(\Ek)-f(\xikk)]\uksq\big\}\nonumber\\
&=& - (n_2 -n_3)\,. \ea

Thus we deduce \be I(\nu) =
-\frac{1}{\pi}\mbox{Im}D^R(\nu)=(n_2-n_3) \delta(\nu)\,, \ee which
is the expected \cite{YuBaym} result.

\section{ Details on Explicit Evaluation of Sum Rules}
\label{App:C}

We will next prove the sum rules on the zeroth and first moment of the RF
spectra for the unpolarized case.
We
focus on the general case $a_{12}\neq a_{13}$. In the large
$\nu$ limit,  from
Eqs.~(\ref{Bf}) and
(\ref{Cf})
it follows that
$B\sim \nu^{-1/2}$ and $C\sim \nu^{-1/2}$. Thus we have
$\chi_{13}(\nu)=\chi_{13}(0)-B\nu\sim \nu^{1/2}$ as $\nu\to\infty$.
We deduce
%which means, as in Reference \cite{Strinati7}
$t_{13}(\nu)=[1/g_{13}+\chi_{13}(\nu)]^{-1}\sim\nu^{-1/2}$. Thus
we have $I_0\sim \nu^{-3/2}$ and $I\sim\nu^{-5/2}$ as
$\nu\to\infty$.
Consequently, we see that
final state effects convert a diverging first
moment into a finite result for the clock shift.

To obtain this shift quantitatively, we use
%moment of RF current finite. Since $I_0\to 0$ as $\nu\to\infty$,
the Kramers-Kronig relations for $t_{13}$. \ba
\mbox{Re}\,t^{R}_{13}(\nu)=-\int^{+\infty}_{-\infty}
\frac{\d\nu'}{\pi}
\frac{\mbox{Im}\,t^{R}_{13}(\nu')}{\nu-\nu'}\label{KK} \ea

For convenience we define $\delta g=\frac{m}{4\pi a_{13}} -
\frac{m}{4\pi a_{12}}$ so that
$$
\mbox{Re}\,t^{R}_{13}(\nu)=\frac{\delta g-B\nu}{(\delta
g-B\nu)^2+(C\nu)^2}
$$

Then from the Kramers-Kronig relations with $\nu=0$ we have
\ba
\int^{+\infty}_{-\infty}\frac{\d\nu'}{\pi}\frac{\mbox{Im}\,t^{R}_{13}(\nu')}{\nu'}
=\mbox{Re}\,t^{R}_{13}(0)=\delta g^{-1} \ea We also have
by taking a derivative and setting
$\nu=0$,  \ba
\int^{+\infty}_{-\infty}\frac{\d\nu'}{\pi}\frac{\mbox{Im}\,t^{R}_{13}(\nu')}{\nu'^2}
=\frac{\p}{\p\nu}\mbox{Re}\,t^{R}_{13}(0)=B(0)/\delta g^2 \ea

From these results we
can prove the sum rule \ba \int d\nu I(\nu)&=&-\delta
g^2\Delta^2\int\frac{d\nu}{\pi}\frac{\mbox{Im}t_{13}^R(\nu)}{\nu^2}\nonumber\\
&=& -B(0) \Delta^2= n_2-n_3 \ea

Then the first moment is \ba \int d\nu \,\nu I(\nu)&=&-\delta
g^2\Delta^2\int\frac{d\nu}{\pi}\frac{\mbox{Im}t_{13}^R(\nu)}{\nu^2}\nonumber\\
&=&-\delta
g\Delta^2=\Delta^2\frac{m}{4\pi}\left(\frac{1}{a_{12}}-\frac{1}{a_{13}}\right)\ea

Combining these equations we find for the mean clock shift
\ba \bar{\nu}=\frac{\int \d\nu \,\nu I(\nu)} {\int
  \d\nu \,I(\nu)}= \frac{ \Delta^2}{n_2 - n_3}
 \frac{m}{4\pi} \left(\frac{1}{a_{12}} - \frac{1}{a_{13}}\right)
\ea
which is the expected result \cite{YuBaym}. In the text we explore how
this sum rule may be used to measure the pairing gap.

\section{ Final State Effects in A Homogeneous But Polarized System}

In the polarized case, state $1$ and $2$ have different chemical
potentials. All the calculations will closely parallel the
unpolarized case. In the following, we only consider the RF
transition from state $2$ to $3$. The formulae for the transition from
from state $1$ to $3$ can be similarly derived. As before, we define
\begin{eqnarray}
F'(K)&=&-\Delta G^{(2)}(K)
G_0^{(1)}(-K)\nonumber\\
&=&-\Delta\frac{1}{i\omega_l-\xik-h}
\frac{-i\omega_l+\xik+h}{(i\omega_l-h)^2-\Ek^2}\nonumber\\
&=&\frac{\Delta}{(i\omega_l-h)^2-\Ek^2}
\end{eqnarray}
Here $\mu=(\mu_{\uparrow}+\mu_{\downarrow})/2$ and
$h=(\mu_{\uparrow} -\mu_{\downarrow})/2$,

Generalizing $D_2(Q)$ to the polarized case we find \ba
D'_2(Q)&=&\sumk \frac{\Delta}{2\Ek}
\bigg[\frac{1-f(\Eku)-f(\xikk)}{(i\Omega_n+h)-\Ek-\xikk}\nonumber\\
&-&\frac{f(\Ekd)-f(\xikk)}{(i\Omega_n+h)
  +\Ek -\xikk}\bigg] \nonumber
\ea with $\Eku=\Ek-h$ and $\Ekd=\Ek+h$.

We analytically continue along with a shift of variables to
write:  \be i\Omega_n
\rightarrow \Omega + i0^+ = \nu +\mu_2 -\mu_3 +i0^+ \quad \ee In this way
we find
the denominator in $D_2(Q)$ is the same as
for the unpolarized case; the only difference
is in the numerator. \ba
1-f(\Ek)-f(\xikk)&\to&1-f(\Eku)-f(\xikk)\nonumber\\
f(\Ek)-f(\xikk)&\to&f(\Ekd)-f(\xikk)\nonumber \ea

We can similarly evaluate $D_0(\nu)$ and $\chi_{13}(\nu)$.
Finally, we not that \ba
\mbox{Im} {D'}_0^{R} &=& \frac{\Delta^2}{\nu}  C_2\\
\mbox{Re}{t'}_{13}^{-1,R}&=&\frac{m}{4\pi a_{13}}-\frac{m}{4\pi a_{12}}-B_2\nu\\
\mbox{Im}{t'}_{13}^{-1,R}&=&-C_2\nu\ea with

\ba B_2(\nu) &=& \sumk \frac{1}{2\Ek} \Big[
\frac{1-f(\Eku)-f(\xikk)}{\nu-\Ek -\xik}
-\frac{f(\Ekd)-f(\xikk)}{\nu  +\Ek -\xik} \Big]\nonumber\\
C_2(\nu) &=& -\pi \sumk \frac{1}{2\Ek} \Big\{ [1-f(\Eku)-f(\xikk)]\delta(\nu-\Ek -\xik)\nonumber\\
&-& [f(\Ekd)-f(\xikk)]\delta(\nu +\Ek -\xik) \Big\}\nonumber \ea

In this way we find for the RF current of state $2$. \ba
I_2(\nu)&=&-\frac{1}{\pi}\left(\frac{m}{4\pi a_{13}} -
\frac{m}{4\pi a_{12}}\right)^2\frac{\mbox{Im}
{D'}_0^R(\nu)}{|{t'}_{13}^{-1,R}(\nu)|^2}\nonumber \ea which is
formally the same result we found in the absence of population
imbalance. These results can readily be generalized to compute the
current in state 1.

\subsection{Sum Rules for the Polarized Case}

Just as for the unpolarized case, we have the Kramers-Kronig
relations (\ref{KK}) for $t_{13}^R$. Following the same analysis
as for the unpolarized case we have
\ba
\int^{+\infty}_{-\infty}\frac{\d\nu'}{\pi}\frac{\mbox{Im}\,{t'}^{R}_{13}(\nu')}{\nu'}
&=&1/\delta g\nonumber\\
\int^{+\infty}_{-\infty}\frac{\d\nu'}{\pi}\frac{\mbox{Im}\,{t'}^{R}_{13}(\nu')}{\nu'^2}
&=&B_2(0)/\delta g^2 \nonumber\ea

Now using the
fact that $\Delta^2B_2(0)=-(n_2-n_3)$, it is straightforward to obtain the zeroth and
first moment of the RF current
\ba \int d\nu I_2(\nu)&=&n_2-n_3\nonumber\\
 \int d\nu \nu\,I_2(\nu)&=&-\delta g\Delta^2 \nonumber\ea
Then the average clock shift is \ba \bar{\nu}=\frac{\int \d\nu
\,\nu I_2(\nu)} {\int
  \d\nu \,I_2(\nu)}= \frac{ \Delta^2}{n_2 - n_3}
 \frac{m}{4\pi} \left(\frac{1}{a_{12}} - \frac{1}{a_{13}}\right)
\nonumber\ea
Again, this is formally, the same as for the unpolarized case.

\vspace*{-1ex}

\bibliographystyle{apsrev}
%\bibliography{Review2}

\begin{thebibliography}{77}
\bibitem[{cav()}]{caveat}
\bibinfo{note}{Above $T_c$ one should ultimately include the effects associated
  with non-zero $\mu_{pair}$. We ignore them here for simplicity and use strict
  mean field theory to estimate the excitation gap everywhere.}
\end{thebibliography}

\end{document}